\documentclass[]{pasj01}
\usepackage{lineno}
\usepackage{url}

\usepackage{color}
\usepackage{ulem}

\draft

\Received{}
\Accepted{}
 
\begin{document} 

\newpage
\title{ 
A Trio of Giant Planets Orbiting Evolved Star HD 184010}

\author{
Huan-Yu \textsc{Teng}\altaffilmark{1,*}, 
Bun'ei \textsc{Sato}\altaffilmark{1},
Takuya \textsc{Takarada}\altaffilmark{2}, 
Masashi \textsc{Omiya}\altaffilmark{2,3}, 
Hiroki \textsc{Harakawa}\altaffilmark{4}, 
Makiko \textsc{Nagasawa}\altaffilmark{5}, 
Ryo \textsc{Hasegawa}\altaffilmark{1}, 
Hideyuki \textsc{Izumiura}\altaffilmark{6}, 
Eiji \textsc{Kambe}\altaffilmark{4}, 
Michitoshi \textsc{Yoshida}\altaffilmark{4}, 
Yoichi \textsc{Itoh}\altaffilmark{7},
Hiroyasu \textsc{Ando}\altaffilmark{3}, 
Eiichiro \textsc{Kokubo}\altaffilmark{8}, and 
Shigeru \textsc{Ida}\altaffilmark{9} 
}

\email{teng.h.aa@m.titech.ac.jp}

\altaffiltext{1}{Department of Earth and Planetary Sciences, School of Science, Tokyo Institute of Technology, 2-12-1 Ookayama, Meguro-ku, Tokyo 152-8551, Japan}
\altaffiltext{2}{Astrobiology Center, National Institutes of Natural Sciences, 2-21-1 Osawa, Mitaka, Tokyo 181-8588, Japan}
\altaffiltext{3}{National Astronomical Observatory of Japan, National Institutes of Natural Sciences, 2-21-1 Osawa, Mitaka, Tokyo 181-8588, Japan}
\altaffiltext{4}{Subaru Telescope, National Astronomical Observatory of Japan, National Institutes of Natural Sciences, 650 North A’ohoku Pl., Hilo, HI, 96720, USA}
\altaffiltext{5}{Department of Physics, Kurume University School of Medicine, 67 Asahi-machi, Kurume, Fukuoka 830-0011, Japan}
\altaffiltext{6}{Okayama Branch Office, Subaru Telescope, National Astronomical Observatory of Japan, National Institutes of Natural Sciences, Kamogata, Asakuchi, Okayama 719-0232, Japan}
\altaffiltext{7}{Nishi-Harima Astronomical Observatory, Center for Astronomy, University of Hyogo, 407-2, Nishigaichi, Sayo, Hyogo 679-5313, Japan}
\altaffiltext{8}{The Graduate University for Advanced Studies
(SOKENDAI), 2-21-1 Osawa, Mitaka, Tokyo 181-8588, Japan}
\altaffiltext{9}{Earth-Life Science Institute, Tokyo Institute of Technology, 2-12-1 Ookayama, Meguro-ku, Tokyo 152-8551, Japan}

\KeyWords{stars: individual: HD 184010 --- planetary systems --- techniques: radial velocities}  

\maketitle
\begin{abstract}
We report the discovery of a triple-giant-planet system around an evolved star HD 184010 (HR 7421, HIP 96016). 
This discovery is based on observations from Okayama Planet Search Program, a precise radial velocity survey, undertaken at Okayama Astrophysical Observatory between 2004 April and 2021 June. 
The star is K0 type and located at beginning of the red-giant branch. 
It has a mass of $1.35_{-0.21}^{+0.19} M_{\odot}$, a radius of $4.86_{-0.49}^{+0.55} R_{\odot}$, and a surface gravity $\log g$ of $3.18_{-0.07}^{+0.08}$. 
The planetary system is composed of three giant planets in a compact configuration:
The planets have minimum masses of $M_{\rm{b}}\sin i = 0.31_{-0.04}^{+0.03} M_{\rm{J}}$, $M_{\rm{c}}\sin i = 0.30_{-0.05}^{+0.04} M_{\rm{J}}$, and $M_{\rm{d}}\sin i = 0.45_{-0.06}^{+0.04} M_{\rm{J}}$, and orbital periods of $P_{\rm{b}}=286.6_{-0.7}^{+2.4}\ \rm{d}$, $P_{\rm{c}}=484.3_{-3.5}^{+5.5}\ \rm{d}$, and $P_{\rm{d}}=836.4_{-8.4}^{+8.4}\ \rm{d}$, respectively, which are derived from a triple Keplerian orbital fit to three sets of radial velocity data.
The ratio of orbital periods are close to $P_{\rm{d}}:P_{\rm{c}}:P_{\rm{b}} \sim 21:12:7$, which means the period ratios between neighboring planets are both lower than $2:1$. 
The dynamical stability analysis reveals that the planets should have near-circular orbits. 
The system could remain stable over 1 Gyr, initialized from co-planar orbits, low eccentricities ($e=0.05$), and planet masses equal to the minimum mass derived from the best-fit circular orbit fitting. Besides, the planets are not likely in mean motion resonance.
HD 184010 system is unique: 
it is the first system discovered to have a highly evolved star ($\log g < 3.5$ cgs) and more than two giant planets all with intermediate orbital periods ($10^2\ \rm{d} < P < 10^3\ \rm{d}$). 
\end{abstract}


\section{Introduction}
\begin{figure}
\begin{center}
\includegraphics[scale=0.35]{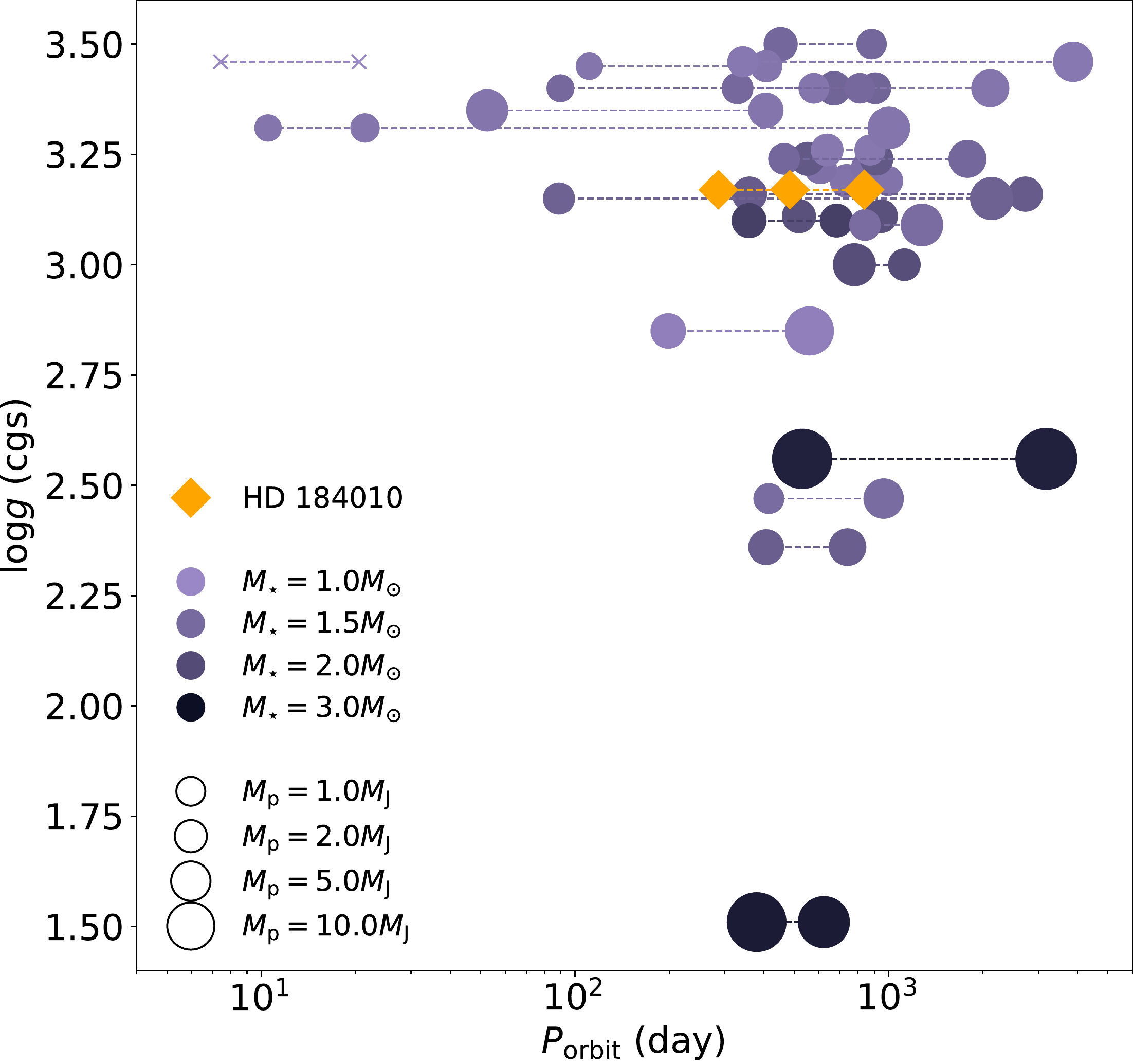} 
\end{center}
\caption{
Multi-planet systems discovered with the radial
velocity and transit methods around evolved stars (26 systems in total). 
We present the distribution of the stellar surface gravity $\log g$ of host star against the orbital period $P_{\rm{orbit}}$ of planet. Each planet is shown by one circle plot, and each individual star is represented by the dashed line connecting its planets. The planet mass (or minimum mass) is shown by the size of the plot, and the star mass is shown by the darkness of the plot. Planets without mass measurements are marked with crosses. 
For HD 184010, planets are in particular marked by orange spades. 
The data except for HD 184010 system were downloaded from NASA Exoplanet Archive.
(A colored version of this figure is available in the online journal.)}\label{fig:logg_prot}
\end{figure}

Nowadays, planets around evolved stars (giants and subgiants) have been extensively surveyed for over 20 years, so as to explore the world of planets around more massive stars, and to investigate how the planetary system evolves after the host star moving out of their main-sequence phase. 
So far, over 150 planet-harboring evolved stars have been confirmed with both radial velocity (RV) method and transit method\footnote{Evolved stars are simply defined as $\log g \lesssim 3.5$, and data acquisition is from NASA Exoplanet Archive \citep{Akeson2013}}. 
Among these systems around evolved stars, only less than 15\% of the systems are known to have more than one planet (shown in Figure \ref{fig:logg_prot}), whereas they could be the key to enigma of evolution of the planetary systems. 
Thanks to long-baseline RV surveys, these multi-planet systems have been discovered.

As shown in Figure \ref{fig:logg_prot},
the majority of these systems are in the pattern of massive planet pairs and have intermediate orbital periods, i.e. $\sim 10^2$ to $\sim 10^3$ days, while the minority includes some hot Jupiters orbiting the host star with short orbital periods.
Anyhow, these planets are likely to have passed Type I, II, or III migration (e.g. \cite{Goldreich1980,Lin1996,Masset2003,Ida2004}), where planets have interactions with the disk, or they may have experienced gravitational interactions to reach the current position, i.e. Kozai mechanism \citep{Wu2003,Fabrycky2007,Nagasawa2008} or planet-planet scattering \citep{Ford2008,Chatterjee2008,Wu2011}.
Besides, many of the planet pairs are considered to be in mean-motion resonance (MMR), and some have period ratio lower than 2:1. 
BD+20 2457 \citep{Niedzielski2009} is a highly evolved ($\log g=1.51$ cgs) K2-giant, orbited by two brown dwarfs with near 3:2 MMR.
HD 33844 \citep{Wittenmyer2016} has two planets in 5:3 resonance. 
7 CMa, HD 5319, HD 99706, HD 102329, HD 116029 and HD 200964 \citep{Wittenmyer2011,Johnson2011a,Giguere2015,Bryan2016,Luque2019} harbor planet pairs with period ratio equal or close to 4:3. 
These discoveries uncovered that massive planets with low period ratios might have passed strong 2:1 MMR during the inward migration \citep{Trifonov2019}. 
Furthermore, HD 33142 \citep{Bryan2016, Trifonov2022} is a newly discovered system centered by an intermediate-mass giant star and has more than two giant planets. It consists of two Jupiter-like planets with orbital periods of $330.0\ \rm{d}$ and $810.2\ \rm{d}$ days, as well as a Saturn-like planets in a close-in orbit with an orbital period of $89.9\ \rm{d}$.
\textit{Kepler}-56 \citep{Huber2013,Otor2016} is another evolved star reported to have more than 2 planets. The inner two co-planar hot Jupiters, which were detected by transit, have 2:1 period ratio, but they are in spin-orbit misaligned. The outermost planet, which was detected by RV follow-up, orbits the star with a period of 1002 days at 2.16 au.

Here we report the RV detection of a triple-giant-planet system around an evolved star HD 184010 (K0 III-IV, $M_{\star} = 1.35 M_{\odot}$, $\log g = 3.18$ cgs), one of the 300 targets in Okayama Planet Search Program (OPSP; \cite{Sato2005}).
It is a unique system since its central star is the first evolved star reported to host more than two intermediate-period giant planets, and both period ratios between the neighboring two planets are smaller than 2:1. This system could be dynamically stable over 100-million years if it is initialized with co-planar, near-circular, and edge-on orbits.    

The rest of the paper is organized as follows. 
In Section \ref{sec:star}, we provide our estimates of stellar parameters of HD 184010. 
In Section \ref{sec:observation}, we describe OAO observations, data reduction of echelle spectra, and methodology in RV measurements. 
In Section \ref{sec:orbfit}, we provide orbital solutions of the HD 184010 planetary system. 
In Section \ref{sec:line}, we analyze line profile of the spectra and chromospheric activity of the star. 
In Section \ref{sec:dyn} we calculate long-term dynamical stability of the planetary system. 
Finally, in Section \ref{sec:discuss}, we summarize our results and briefly discuss the HD 184010 planetary system. 

\section{Stellar properties}\label{sec:star}
\begin{table}
\tbl{Stellar Properties of HD 184010}{
\begin{tabular}{lrr}
\hline\hline
 & HD 184010 & Source \\ 
\hline 
$\pi\ (\rm{mas})$& $16.2940$ &\textit{Gaia} EDR3  \\ 
$V$& $5.89$  & \textit{Hipparcos}  \\ 
$B-V$& $0.918$  & \textit{Hipparcos}  \\ 
Spec. type & K0III-IV & \textit{Hipparcos}  \\ 
$T_{\mathrm{eff,sp}}\ ({\mathrm{K}})$ & $5011$  & \citet{Takeda2008}  \\ 
$\mathrm{[Fe/H]}_{\mathrm{sp}} \mathrm{(dex)}$ & $-0.14$  & \citet{Takeda2008}   \\ 
$\log g_{\star,sp}\ ({\mathrm{cgs}})$ & $3.17$  & \citet{Takeda2008}  \\ 
$v\sin i (\mathrm{km}\ \mathrm{s}^{-1})$ & $1.34$  & \citet{Takeda2008}   \\  
$T_{\mathrm{eff}}\ ({\mathrm{K}})$ & $4971$  & This work  \\ 
${\mathrm{[Fe/H]\ (dex)}}$ & $-0.17_{-0.10}^{+0.10}$  & This work  \\ 
$\log g_{\star}\ ({\mathrm{cgs}})$ & $3.18_{-0.07}^{+0.08}$  & This work  \\ 
$L_{\star}\ (L_{\odot})$ & $13.09_{-2.65}^{+3.15}$  & This work  \\ 
$M_{\star}\ (M_{\odot})$ & $1.35_{-0.21}^{+0.19}$  & This work  \\ 
$R_{\star}\ (R_{\odot})$ & $4.86_{-0.49}^{+0.55}$  & This work  \\ 
Age (Gyr) & $2.76_{-0.95}^{+2.24}$  & This work  \\ 
\hline
\end{tabular}}
\label{tab:properties}
\end{table}
\begin{figure}
\begin{center}
\includegraphics[scale=0.5]{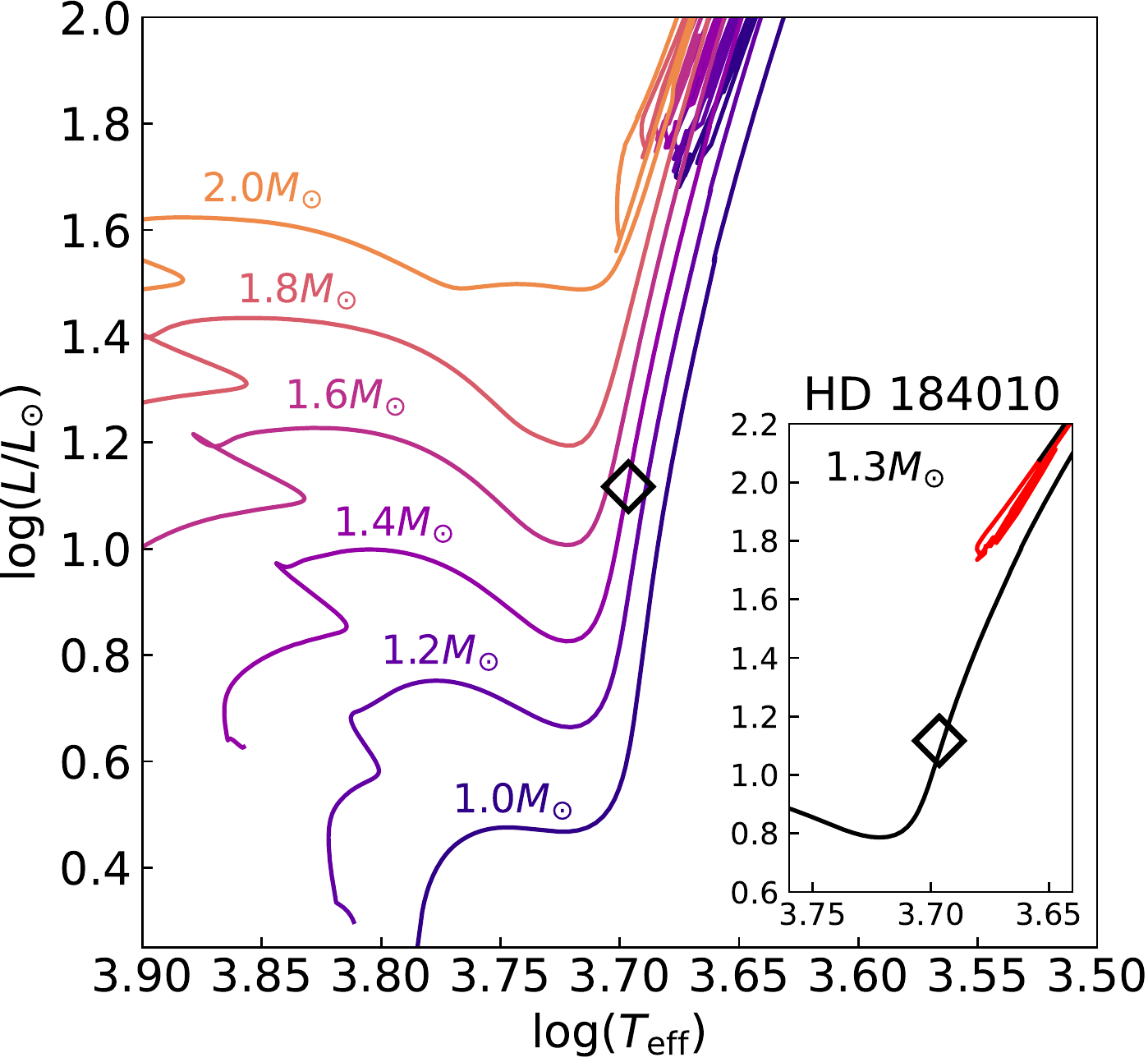} 
\end{center}
\caption{
HR diagram for HD 184010. The star is marked by a hollow spade. Evolution tracks of stars having masses between 1.0 and 2.0 $M_{\odot}$ with HD 184010 metallicity ($\mathrm{[Fe/H]} = -0.17$) are shown in solid lines using different colors in the main figure. Another zoomed figure is given to illustrate the evolution track of stars of 1.3 $M_{\odot}$ with Helium burning stage specially marked in red color.
(A colored version of this figure is available in the online journal.)
}\label{fig:HR}
\end{figure}
\begin{figure}
\begin{center}
\includegraphics[scale=0.4]{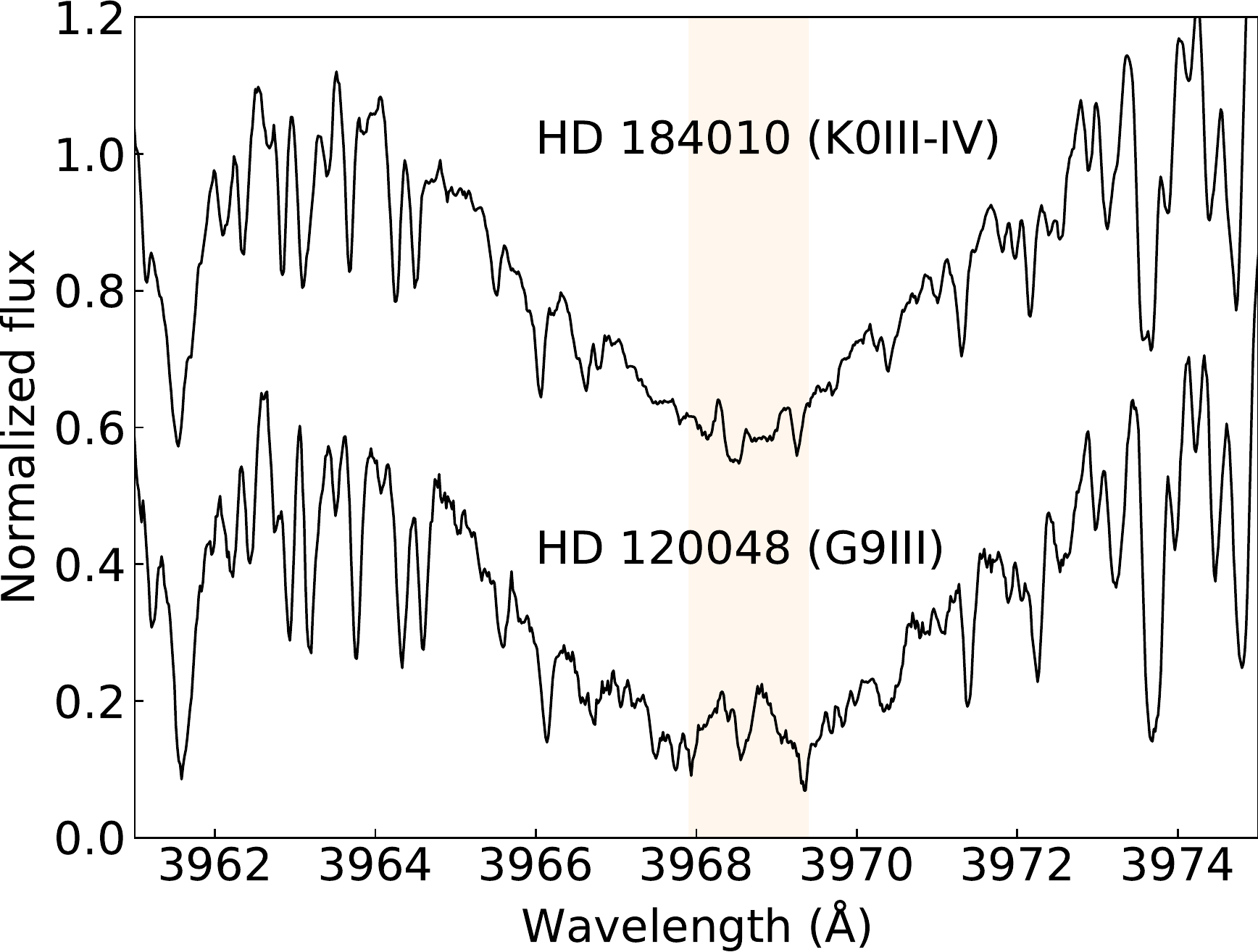}
\end{center}
\caption{
Spectra in the region of Ca \emissiontype{II} H lines of HD 184010 with an active star HD 120048 as a comparison. 
Vertical offsets are added to each normalized spectrum for clarity, and a shaded area is used to mark the core region.
}\label{fig:CaIIH}
\end{figure}

HD 184010 (HR 7421, HIP 96016) is listed in the \textit{Hipparcos} catalog \citep{ESA1997} as K0 III-IV star with apparent magnitude $V$-band $V=5.89$, and its parallax is obtained from \textit{Gaia} EDR3 \citep{Gaia2016,Gaia2021,Lindegren2021} with $\pi = 16.2940 \pm 0.0301$ mas, corresponding to a distance of $61.3722_{-0.1114}^{+0.1119}$ pc. The atmospheric parameters (effective temperature $T_{\rm{eff,sp}}$, surface gravity $\log g_{\rm{sp}}$, and Fe abundance [Fe/H]$_{\rm{sp}}$) were determined by \citet{Takeda2008} by measuring equivalent width of Fe \emissiontype{I} and Fe \emissiontype{II} lines of iodine-free stellar spectra. They also determined projected rotational velocity, $v\sin i$ with the automatic spectrum-fitting technique \citep{Takeda1995}. For star HD 184010, they obtained $T_{\mathrm{eff,sp}}=5011\  {\mathrm{K}}$, $\mathrm{[Fe/H]}_{\mathrm{sp}} = -0.14\ \mathrm{(dex)}$, $\log g_{\star,sp} = 3.17\ ({\mathrm{cgs}})$ as well as $v\sin i = 1.34\ \mathrm{km}\ \mathrm{s}^{-1}$.

\citet{Takeda2008} also estimated the stellar mass of HD 184010 to be 1.82 $M_{\odot}$, with luminosity they determined, effective temperature $T_{\mathrm{eff,sp}}=5011\  {\mathrm{K}}$, and theoretical evolutionary tracks from \citet{Lejeune2001}. 
However, according to a work given by \citet{Takeda2015}, stellar masses in \citet{Takeda2008} tend to be overestimated by up to a factor of $\lesssim 2$, especially for giant stars located near the red clump regions in the Hertzsprung-Russell (HR) diagram. 
This is mainly on account of the coarse evolutionary tracks under a low-resolution parameter grid and deficiency of ``He flash'' red-clump-giant for stars less-massive than $2 M_{\odot}$. Fine tracks under high-resolution parameter grid could mitigate the over-estimation.

For this reason, we re-estimated stellar parameters with the same method in \citet{Teng2022} by adopting the Bayesian estimation method with theoretical isochrones with \texttt{isoclassify} \citep{Huber2017, Berger2020}.
We used the \texttt{isoclassify} ``direct'' mode and applied the Stefan-Boltzmann law to derive $R_{\star}$ and $L_{\star}$ from posterior probability distributions with spectroscopic $T_{\mathrm{eff,sp}}$ and $V$-band photometry as inputs. In this step, we sampled the distance following the parallax posterior from \textit{Gaia} EDR3. For each distance sample, we calculate the extinction $A_{V}$ with map given by \citet{Green2019} as implemented in \texttt{mwdust} package by \citet{Bovy2016}. Besides, each sample was combined with independent random normal samples for apparent magnitude and effective temperature $T_{\mathrm{eff}}$. We derive the bolometric corrections by linearly interpolating $T_{\mathrm{eff}}$, $\mathrm{[Fe/H]}$, $\log g_{\star}$, and $A_{V}$ in the \texttt{MIST}/C3K grid \citep{ConroyInPrep}\footnote{(\url{http://waps.cfa.harvard.edu/MIST/model_grids.html})}. We iterated distances, extinctions, and bolometric corrections until their convergence. 
Consequently, we obtained $R_{\star} = 4.86_{-0.49}^{+0.55} R_{\odot}$ and $L_{\star}= 13.09_{-2.65}^{+3.15} L_{\odot}$ for HD 184010.
Next, we used the \texttt{isoclassify} ``grid'' mode to estimate parameters including $T_{\mathrm{eff}}$, $\mathrm{[Fe/H]}$, $\log g_{\star}$, $L_{\star}$, $R_{\star}$, and $M_{\star}$ with posterior distribution by integrating over \texttt{MIST} isochrone \citep{Paxton2011, Paxton2013, Paxton2015, Choi2016, Dotter2016, Paxton2018}. Consequently, we obtain the mass $M_{\star}=1.35_{-0.21}^{+0.19} M_{\odot}$, which is lower than the estimation in \citet{Takeda2008}. The properties of the HD 184010 are listed in table \ref{tab:properties} and plotted on the HR diagram in Figure \ref{fig:HR}.

HD 184010 is stable in $V$-band photometry of different surveys: 1) \textit{Hipparcos} $\sigma_{\rm{HIP}} = 0.006\ \rm{mag}$ (ESA: \cite{vanLeeuwen2007}); 2) All Sky Automated Survey (ASAS-3) $\sigma_{\rm{ASAS-3}} = 0.04\ \rm{mag}$ \citep{Pojmanski1997}; 3) All-Sky Automated Survey for Supernovae (ASAS-SN) $\sigma_{\rm{ASAS-SN}} = 0.05\ \rm{mag}$ \citep{Shappee2014, Kochanek2017}. Detrending was performed to ASAS-3 and ASAS-SN light curves.
Here we also convey the warning from ASAS-SN database that the exposure of HD 184010 could be saturated.
It is chromospherically inactive with no significant emission in the core of the Ca \emissiontype{II} H lines (Figure \ref{fig:CaIIH}). We use another chromospherically active G-type giant star (HD 120048) as a comparison, and together show two spectra in Figure \ref{fig:CaIIH}, where we can see clear emission lines in the core of the Ca \emissiontype{II} H lines in HD 120048 spectrum.

\section{Observations and RV measurements}\label{sec:observation}
In this work, we obtained the spectra of HD 184010 from 1.88-m reflector with HIgh Dispersion Echelle Spectrograph (HIDES: \cite{Izumiura1999}) at Okayama Astrophysical Observatory. We obtained its first spectrum in April 2004 under the Okayama Planet Search Program \citep{Sato2005}, a large planet survey focusing on RV measurements to late-G (including early-K) giant stars, and aiming at discovering planets planets around intermediate-mass stars.

Here we briefly review the HIDES at OAO before 2018 and introduce the latest upgrade after that. The instrument includes an iodine cell placed in the HIDES optical path, which provided numerous iodine absorption lines in the range of 5000-5800 \AA ~as a reference for precise radial velocity measurements. To cover these iodine absorption lines, the wavelength region of HIDES was firstly set to cover 5000-6100 \AA ~with one $2 \mathrm{K} \times 4 \mathrm{K}$ CCD. In December 2007, the CCD of HIDES was upgraded from the single one to a mosaic of three, which widened the wavelength region to 3700-7500 \AA ~(3700-5000 \AA, ~5000-5800 \AA ~and ~5800-7500 \AA ~for each respectively). The upgrade enabled us to simultaneously measure the level of stellar activities (e.g. Ca \emissiontype{II} HK lines) and line profiles as well as radial velocities.

In 2010, a new high-efficiency fiber-link system with its own iodine cell was installed to the HIDES which greatly enhanced the overall throughput \citep{Kambe2013}. 
In 2018, another upgrade was carried out to enhance the performance of fiber-link system\footnote{A more detailed introduction to the HIDES fiber-link mode upgrade and its performance will be given in a forthcoming paper.}. In this upgrade, the slit mode elements were fully removed from the HIDES optical path, and optical instruments of fiber-link mode were re-arranged on a new stabilized platform in the precise temperature-controlled Coud\'e room. The upgrade finished in 2018 December, and after that HIDES observation restarted.
Since the optical path was re-arranged in the upgrade, fiber-link mode before/after the upgrade are considered to be two independent instruments. 

In this research, the spectra were obtained by both conventional slit mode (hereafter HIDES-S) and fiber mode (pre-upgrade in 2018: HIDES-F1, and post-upgrade in 2018: HIDES-F2). 

In the case of HIDES-S observations, the slit width was set to 200 $\mu$m ($\timeform{0.76''}$) corresponding to the resolution $\mathit{R} = \lambda / \Delta \lambda \sim 67000$ by about 3.3-pixel sampling. In the case of HIDES-F observations, the width of the sliced image was $\timeform{1.05''}$ corresponding to the resolution $\mathit{R} \sim 55000$ by about 3.8-pixel sampling. There could be an offset in radial velocity measurements between two modes due to their respective iodine in their optical path, hence we treated two modes as two different instruments in our research. We adopted the data which gained signal-to-noise ratio (S/N) over approximately 100 per pixel at $\sim$5500 \AA ~within 1800 seconds. For HD 184010, the typical exposure time and S/N is 1200 seconds and over 150, 600 seconds over 200, in the case of slit mode (HIDES-S) and fiber mode (HIDES-F1, and -F2) observations, respectively.

The reduction of these echelle spectra was performed with IRAF\footnote{IRAF \citep{{Tody1986}} is distributed by the National Optical Astronomy Observatories, which is operated by the Association of Universities for Research in Astronomy, Inc. under a cooperative agreement with the National Science Foundation, USA} packages in a standard way (i.e. bias subtraction, flat fielding, scattered light subtraction, and spectrum extraction). Especially, for spectra obtained by fiber-mode, there were severe aperture overlaps among 3700--4000 \AA ~owing to the use of an image slicer, therefore the scattered light for these apertures could not be subtracted with IRAF task $\mathtt{apscatter}$. Consequently, we discard these overlapped apertures and did not analyze the Ca \emissiontype{II} H lines for fiber mode (HIDES-F1 and -F2) spectra.

For precise RV measurements, we used spectra covering 5000 to 5800 \AA ~in which I$_{\rm{2}}$ absorption lines are superimposed by an I$_{\rm{2}}$ cell. We computed RV variations following the method described in \citet{Sato2002} and \citet{Sato2012} which was based on a method by \citet{Butler1996a}. 
We modeled the spectra (I$_{\rm{2}}$ superposed stellar spectra) by using the stellar template which were obtained by deconvolving pure stellar spectra with the IP estimated from I$_{\rm{2}}$ superposed B type star or flat spectra. Hundreds of segments at a typical width of 150 pixels were set in the observed spectra in the I$_{\rm{2}}$ absorption region, and the final RV values and their measurement errors were taken from the average of measurement in each segment. 
Since HIDES-S, -F1, and -F2 have different optical paths, RV offsets between different instruments were given as free parameters in orbit fitting.

\section{Orbit fitting and planetary parameters}\label{sec:orbfit}
\begin{figure}
\begin{center}
\includegraphics[scale=0.45]{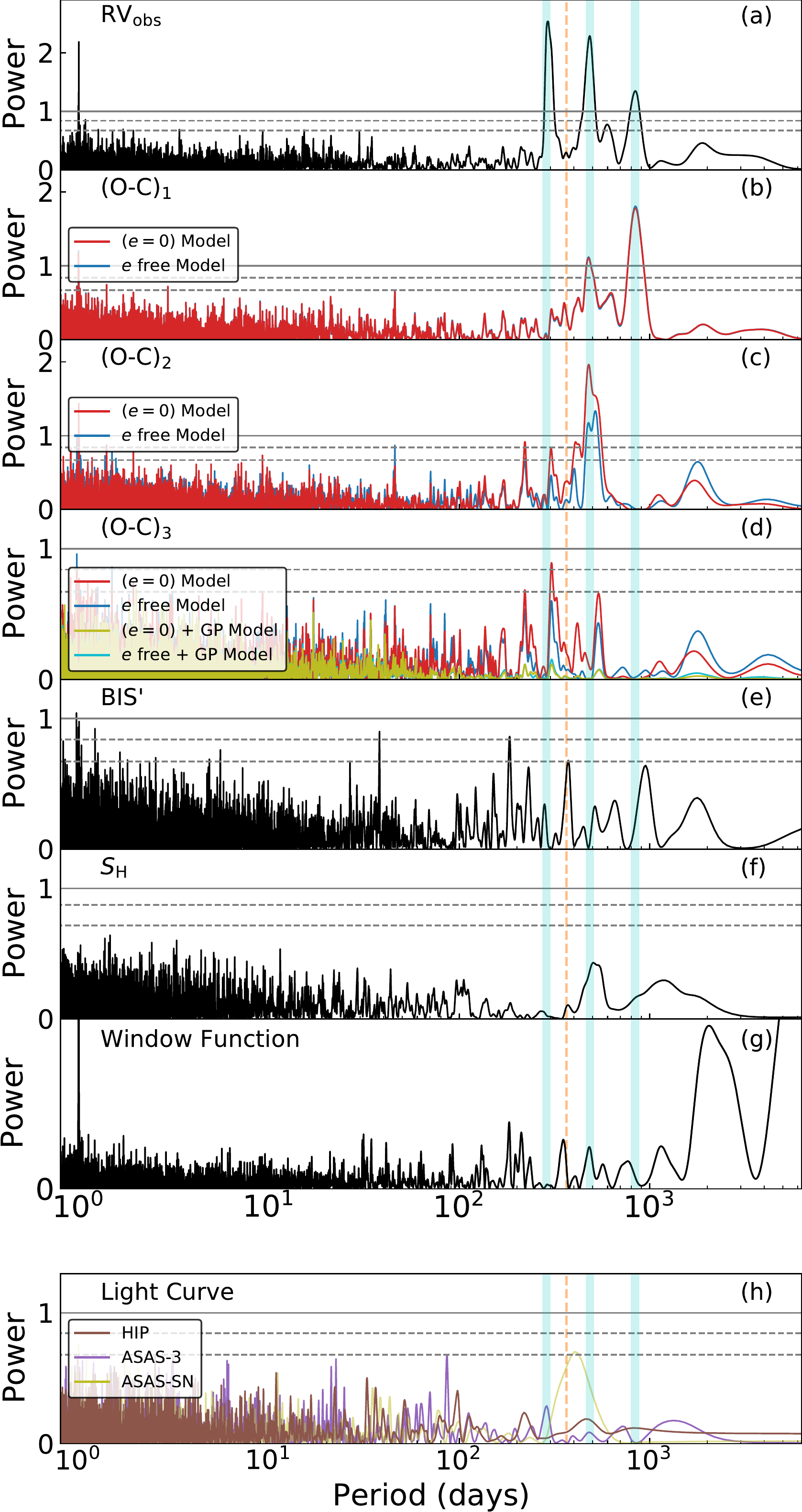} 
\end{center}
\caption{
Generalized Lomb-Scargle (GLS) periodograms for star HD 184010. In all subplots, vertical axes are given in power normalized by 1) methodology given in \citep{Baluev2008}; 2) the power of False Alarm Probability (FAP) equal to 0.1\%. The horizontal axes are given in periods (days). 
Subplot (a) is the GLS periodograms of the observed raw RVs (RV$_{\rm{obs}}$).
Subplot (b, c, d) are the GLS periodograms of residuals to 1-, 2-, and 3-Keplerian fittings [(O-C)$_{\rm{1}}$, (O-C)$_{\rm{2}}$, and (O-C)$_{\rm{3}}$]; 
The red, blue, yellow, and cyan curves indicate fittings based on $(e=0)$ Model, $e$ free Model, $(e=0)$ $+$ GP Model, and $e$ free $+$ GP Model respectively.
Subplot (e, f) are the GLS periodograms of mean removed BIS (BIS') and Index of Ca \emissiontype{H} lines ($S_{\rm{H}}$), respectively.
Subplot (g) is the window function of OAO observations.
Subplot (h) is the GLS periodograms of light curves of three $V$-band light curves. 
The brown, purple, and yellow curve represent \textit{Hipparcos}, ASAS-3, and ASAS-SN, respectively. But ASAS-SN database warned that the exposure of HD 184010 could be saturated.
The horizontal lines represent 10\%, 1\%, and 0.1\% FAP levels from bottom to top, and particularly the 0.1\% FAP level is in solid style. The vertical cyan solid lines indicate the best-fitted periods from the 3-Keplerian model, and the vertical orange dashed line indicates 1 year. 
 (A colored version of this figure is available in the online journal.)}\label{fig:rvls}
\end{figure}

\begin{figure}
\begin{center}
\includegraphics[scale=0.45]{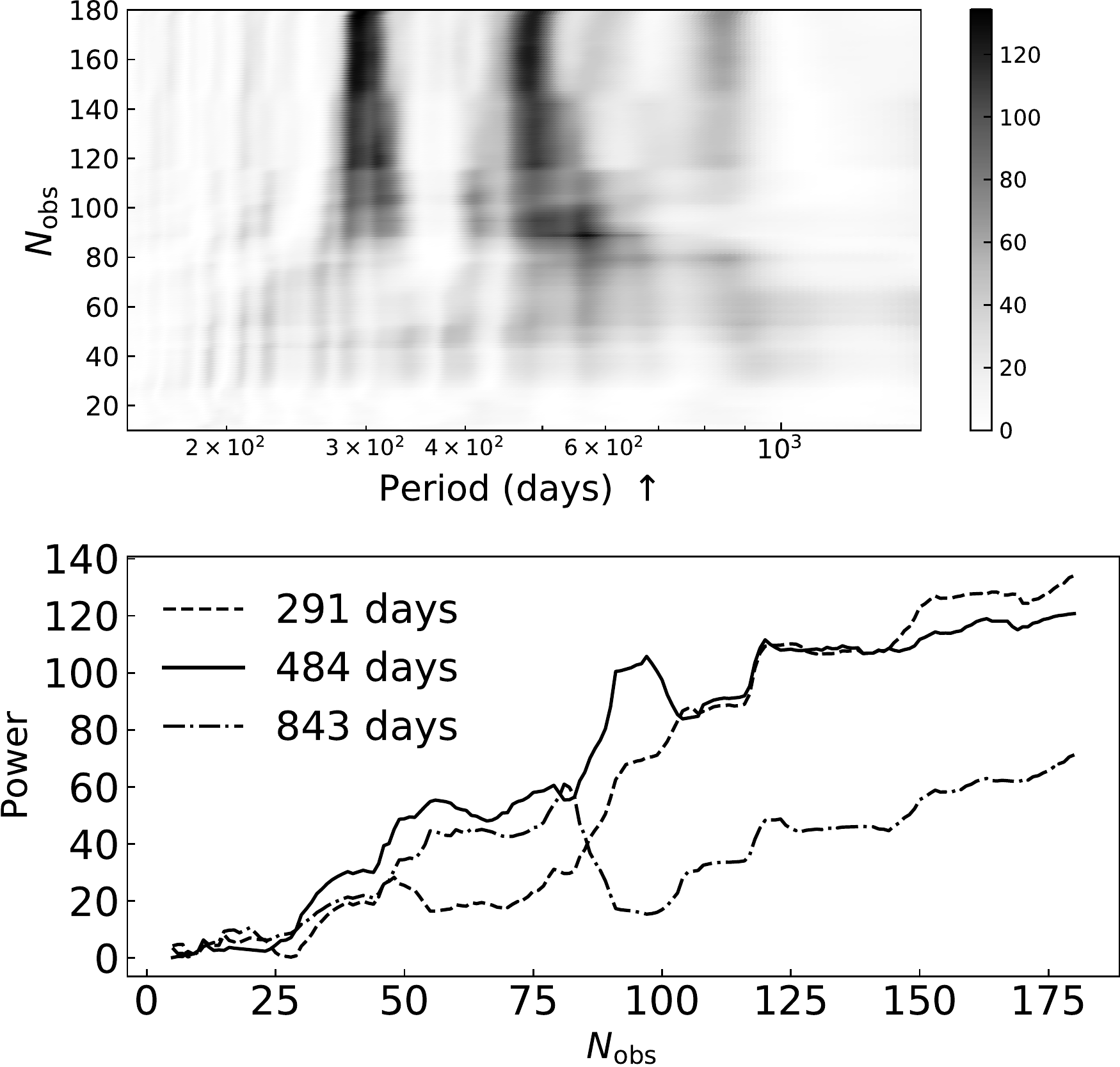} 
\end{center}
\caption{Stacked Bayesian Generalized Lomb-Scargle periodogram of the observed raw RV data of HD 184010. 
The upper panel shows the power variation by period and the number of data points, and the color bar illustrates the strength of power. 
The lower panel shows the power of interested periods varying with the number of data. The legend of 291 days marked with a dashed line, the legend of 484 days marked with a solid line, and the legend of 843 days marked with a dashed-dotted line respectively represent the interested periods of 291-day signal, 484-day, and 843-day signal in the GLS periodogram for observed raw RV data.}\label{fig:rvsbgls}
\end{figure}

In order to filter the high-frequency signals in the time series, we firstly binned the data each night if there were more at least two exposures, and we obtained 39, 81, and 60 data points for HIDES-S, F1, and F2 respectively.
We then investigated the periodicity in the RV time series by performing a Generalized Lomb-Scargle periodogram (GLS: \cite{Zechmeister2009}) with \texttt{astropy} software \citep{Astropy2018}. 
We assessed the significance of the periodicity by applying False Alarm Probability (FAP) with approximation method developed by \citet{Baluev2008} in the same package. 
By performing GLS, we found three significant signals respectively at 291 days, 484 days, and 843 days in the observed raw RVs, and we ruled out the possibility of an alias for all three signals (Figure \ref{fig:rvls}). 
We also employed a Stacked Bayesian Generalized Lomb-Scargle periodogram (SBGLS: \cite{Mortier2017}, Figure \ref{fig:rvsbgls}) to further investigate the evolution of these signals with the increasing amount of data points, and we found the power of these signals all steadily increasing after 100 observations.


Planets' orbital motions can be a self-consistent interpretation for the stable regular variations.
Thus we used 1-, 2- and 3-Keplerian models to fit the three regular RV variations in the sequence of power strength demonstrated by the GLS periodogram.
We tested models of both eccentricity $e$ fixed to 0 (hereafter, $(e=0)$ Model) and $e$ as a free parameter (hereafter, $e$ free Model), since an eccentric orbit could be mimicked by two or more circular orbits.
We also performed GLS periodograms of the residuals to examine if the signals of regular variations were subtracted by Keplerian models. The orbital fits were performed in the following way.

We first generated Keplerian models by $\mathtt{RadVel}$ software \citep{Fulton2017, Fulton2018}, and we adopted orbital parameter set including: orbital period $P$, RV semi-amplitude $K$, the combination of eccentricity $e$ and argument of periastron $\omega$, $\sqrt{e} \cos \omega$ and $\sqrt{e} \sin \omega$, and time of inferior conjunction $T_{\rm{c}}$\footnote{The pariastron passage $T_{\rm{p}}$ is converted from inferior conjunction $T_{\rm{c}}$ after parameter fitting.}. 
We also set RV offsets $\gamma_{\rm{inst}}$ and extra Gaussian noises $s_{\rm{inst}}$ of different instruments as free parameters and fit them together with orbital parameters.

To obtain the best-fit parameters of the Keplerian model, we maximized the likelihood function using a truncated Newton (\textsf{TNC}) algorithm and a \textsf{Nelder-Mead} algorithm \citep{Nelder1965}.
For parameter distribution and uncertainty analysis, we performed Markov Chain Monte Carlo MCMC) sampling with $\mathtt{emcee}$ package\citep{Foreman-Mackey2013} from the maximum likelihood results.
We set priors to the free parameters, including Jeffrey's prior for $P$, uniform distribution and $\gamma_{\rm{inst}}$, and modified Jeffrey's prior for $K$ and $s_{\rm{inst}}$. The specific values for each prior are summarized in Table \ref{tab:prior}. 
We generate 5000 MCMC steps per walker per CPU core and run 100 walkers on each of 8 CPU cores. 
The first 1000 MCMC steps in each walker were discarded from the analysis for burn-in in a standard scheme. 
Our MCMC chains had mean acceptance rates between $\sim 0.2$ to $\sim 0.4$, the Gelman-Rubin statistic levels of GR $<1.01$ \citep{Gelman2013}, and independent samples $T_z$ greater than 1000 \citep{Ford2006}. 

Consequently, we obtained final best-fit orbital solutions from the posterior with 1-, 2-, and 3-Keplerian models. 
The periodograms of the residuals showed that the 291-day, 843-day, and 484-day signals could be subtracted in turn by 1-, 2-, and 3-Keplerian models of both circular and eccentric orbits.
There was no significant variability remained in the residuals under 3-Keplerian fits. 

In addition, correlated noise component, known as ``red noise'', is common in  RV time series. 
\citet{Baluev2013} applied a Gaussian process model (GP) to study the impact of red noise in RV planet searches.
To further characterize the prospective red noise caused by the instrument or the star, we also introduced GP to orbit fitting with a squared exponential kernel provided by \texttt{RadVel}. 
The covariance matrix is:
\begin{eqnarray}
C_{\mathrm{ij}} = \eta_1^2  \exp{(\frac{ -|t_\mathrm{i} - t_\mathrm{j}|^2 }{ \eta_2^2 })}.
\end{eqnarray}
We tested GP regression on both 3-Keplerian plus $(e=0)$ and 3-Keplerian plus $e$ free Model, and we derived the orbital solutions including GP hyper parameters ($\eta_1$ and $\eta_2$) in the aforementioned way.

\begin{figure}
\begin{center}
\includegraphics[scale=0.32]{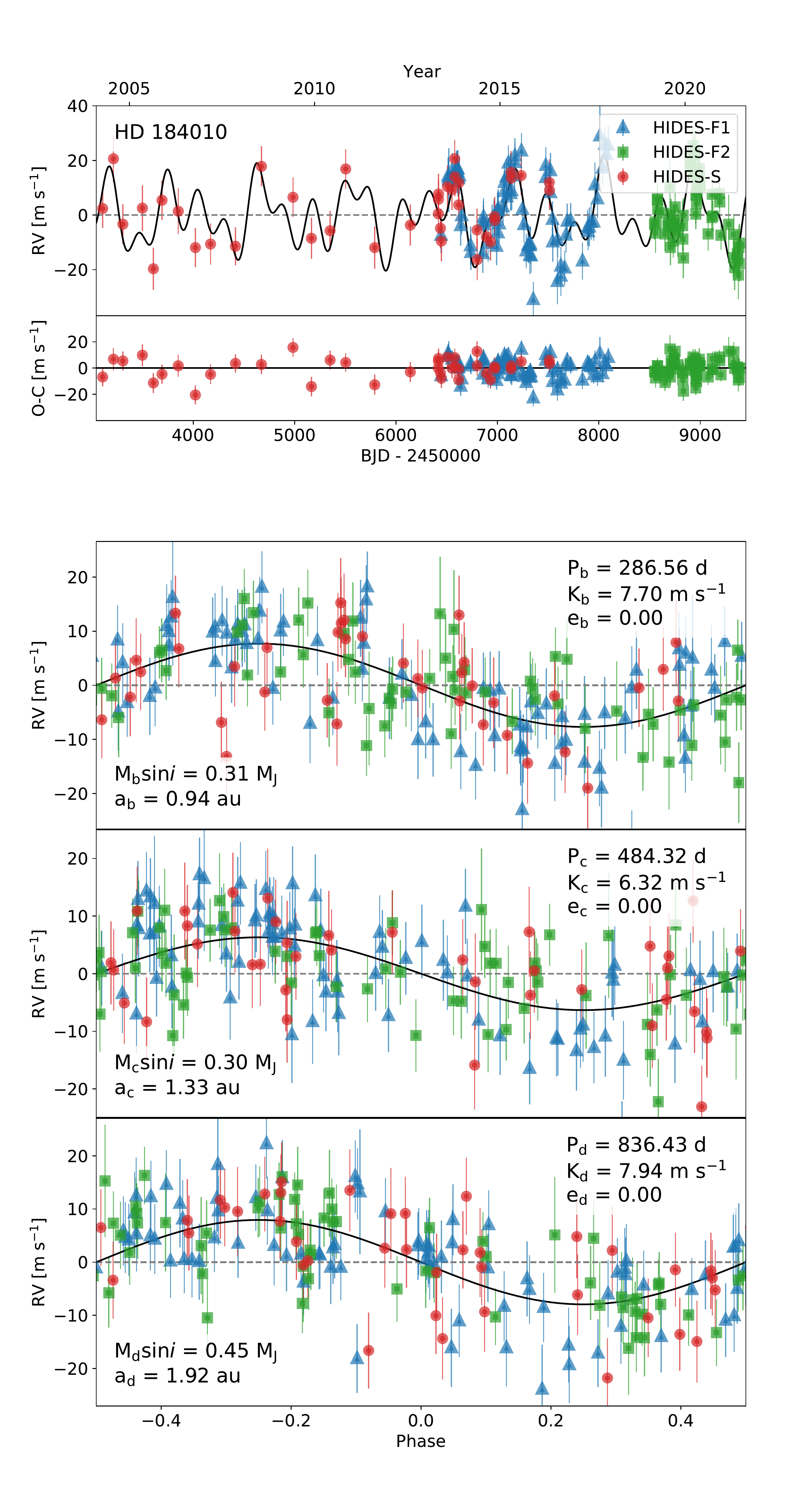} 
\end{center}
\caption{Orbital solution of HD 184010. Subplots from top to bottom: 
Panel (1) shows a general view of the orbit. The black solid line shows the best-fit 3-Keplerian curve along the full-time-span RV time series. The colored dots with errorbars are RVs with fitted RV offsets shifted between instruments and fitted jitters quadratically added to the observational errorbars. 
Panel (2) shows RV residuals to the best-fit 3-Keplerian model. The rms scatter is $7.0\ \rm{m\ s^{-1}}$. 
Panel (3,4,5) from top to bottom are phase-folded orbits of planet b, c, and d, respectively. 
In all subplots, the symbols are the same. HIDES-S data are shown in red circles, HIDES-F1 data are shown in blue triangles, and HIDES-F2 data are shown in green squares. 
(A colored version of this figure is available in the online journal, and a complete RV data listing
will be available online as supplementary after the publication.)}\label{fig:rv}
\end{figure}

\begin{figure*}
\begin{center}
\includegraphics[scale=0.36]{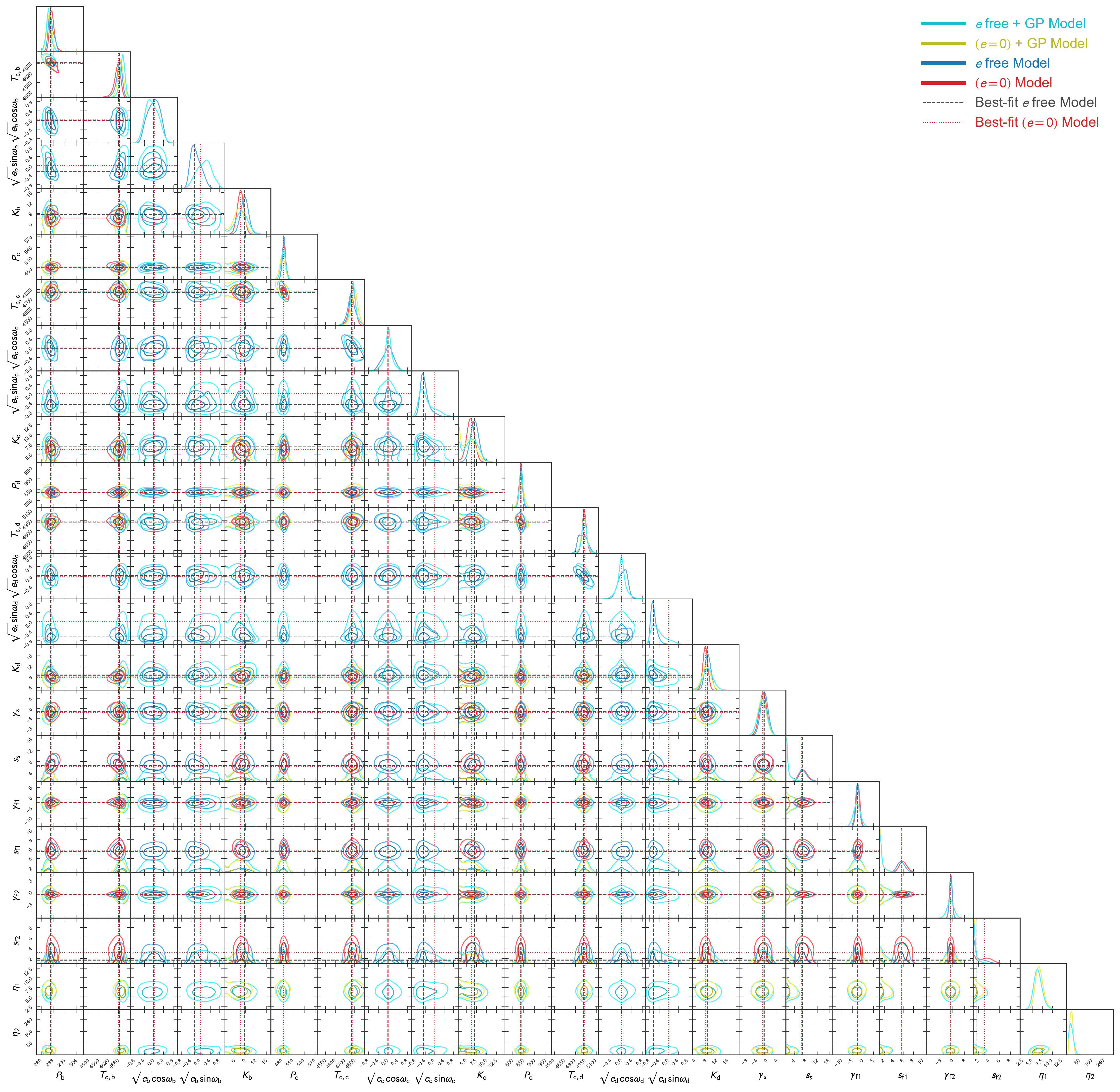} 
\end{center}
\caption{
Contour plots showing MCMC posterior distributions of fitted parameters, 
including orbital period $P$, time of inferior conjunction $T_{\rm{c}}$, combination of eccentricity and longitude of periastron passage $\sqrt{e}\cos \omega$ and $\sqrt{e}\sin \omega$, RV semi-amplitude $K$, instrumental RV offset $\gamma$, and extra RV jitter $s$, and hyper parameters of squared exponential kernel (\texttt{RadVel}: $\eta_{1}$, $\eta_{2}$) of Gaussian Process (GP).
The lower subscript ``s'', ``f1'', and ``f2'' respectively refer to HIDES-S, -F1, and -F2.
We use different colors to illustrate different models in the fittings.
Red: $(e=0)$ Model; Blue: $e$ free Model;
Yellow: $(e=0)$ $+$ GP Model; Cyan: $e$ free $+$ GP Model.
We use straight lines to illustrate the posterior medians of models without GP.
The red dotted lines and grey dashed lines are the best-fit $e=0$ Model and $e$ free Model, respectively.
(A colored version of this figure is available in the online journal.)}\label{fig:corner}
\end{figure*}

We report the MCMC posterior medians and 1$\sigma$ credible regions of four different 3-Keplerian models in Table \ref{tab:orbpar}.
Specifically, we show the final best-fit $(e=0)$ Model in Figure \ref{fig:rv}. 
We also display the posterior distributions of the four models in Figure \ref{fig:corner}.
In our subsequent analyses, we make comparisons between four different 3-Keplerian models and other models, and we also make comparisons within these four models. 
\begin{itemize}
    \item $(e=0)$ Model,
    \item $e$ free Model,
    \item $(e=0)$ $+$ GP Model,
    \item $e$ free $+$ GP Model.
\end{itemize}

\begin{table*}
\tbl{Priors for MCMC sampling.}{%
\begin{tabular}{lcccc}
\hline\hline
Parameter & Prior & Minimum & Maximum & knee value \\ 
\hline
$P\ (\rm{d})$ & Jeffrey's & 1 & 5000 & --\\
$K\ (\rm{m\>s^{-1}})$ & Modified Jeffrey's & 1.01 & 1000 & 1 \\
$\sqrt{e} \cos \omega$ & Uniform & -1 & 1 & -- \\
$\sqrt{e} \sin \omega$ & Uniform & -1 & 1 & -- \\
$\gamma_{\rm{inst}}$ & Uniform & -500 & 500 & -- \\
$s_{\rm{inst}}$ & Modified Jeffrey's & 1.01 & 100 & 1 \\
\hline
\end{tabular}}
\begin{tabnote}
\hangindent6pt\noindent
\hbox to6pt{\footnotemark[$*$]\hss}\unskip%
The subscript ``inst'' refers to a certain instrument.
\end{tabnote}
\label{tab:prior}
\end{table*}

\begin{table*}
\tbl{
Posterior estimates of 3-Keplerian models for HD 184010
}{
\scriptsize
\begin{tabular}{lccccr}
\hline\hline
& \textbf{$(e=0)$ Model} & $e$ free Model & $(e=0)$ + GP Model & $e$ free + GP Model &  \\
\hline 
\textbf{Planet b} \\ 
$P\ (\rm{d})$  & $286.6_{-0.7}^{+2.4}$ & $286.4_{-1.3}^{+2.3}$ & $285.4_{-1.3}^{+1.3}$ & $285.1_{-1.5}^{+1.2}$ & fitted \\ 
$K\ (\rm{m\ s^{-1}})$ & $7.70_{-0.96}^{+0.78}$ & $8.74_{-1.05}^{+1.01}$ & $7.15_{-1.63}^{+1.47}$ & $7.89_{-1.79}^{+1.68}$ & fitted \\ 
$\sqrt{e}\cos \omega$ & 0 (fixed) & $-0.019_{-0.277}^{+0.262}$ & 0 (fixed) & $-0.103_{-0.246}^{+0.300}$ & fitted \\ 
$\sqrt{e}\sin \omega$ & 0 (fixed) & $-0.246_{-0.163}^{+0.259}$ & 0 (fixed) & $0.161_{-0.341}^{+0.287}$ & fitted \\ 
$T_{\rm{c}} $(BJD$-2450000$) & $4609.2_{-28.7}^{+7.4}$ & $4685.3_{-18.5}^{+12.2}$ & $4692.9_{-12.8}^{+13.2}$ & $4705.6_{-15.1}^{+8.3}$ & fitted \\ 
$e$ & 0 (fixed) & $0.061_{-0.018}^{+0.195}$ & 0 (fixed) & $0.037_{-0.012}^{+0.283}$ & derived \\ 
$\omega\ (^{\circ})$ & 0 (fixed) & $-94.3_{-43.6}^{+101.2}$ & 0 (fixed) & $122.6_{-246.2}^{+15.1}$ & derived \\ 
$T_{\rm{p}} $(BJD$-2450000$) & $4608.9_{-29.3}^{+7.5}$ & $4824.6_{-255.9}^{+32.4}$ & $4621.6_{-13.1}^{+13.5}$ & $4729.7_{-72.1}^{+43.3}$ & derived \\ 
$M_{\rm{p}}\sin i\ (M_{\rm{J}})$ & $0.31_{-0.04}^{+0.03}$ & $0.34_{-0.04}^{+0.04}$ & $0.28_{-0.06}^{+0.06}$ & $0.30_{-0.07}^{+0.07}$ & derived \\ 
$a\ (\rm{au})$ & $0.940_{-0.001}^{+0.005}$ & $0.940_{-0.003}^{+0.005}$ & $0.938_{-0.003}^{+0.003}$ & $0.937_{-0.003}^{+0.003}$ & derived \\ 
\hline 
\textbf{Planet c} \\ 
$P\ (\rm{d})$  & $484.3_{-3.5}^{+5.5}$ & $484.9_{-4.1}^{+3.5}$ & $480.6_{-5.0}^{+5.5}$ & $483.3_{-4.0}^{+3.2}$ & fitted \\ 
$K\ (\rm{m\ s^{-1}})$ & $6.32_{-1.06}^{+0.81}$ & $7.07_{-1.04}^{+1.03}$ & $5.51_{-2.08}^{+1.71}$ & $6.21_{-2.11}^{+1.69}$ & fitted \\ 
$\sqrt{e}\cos \omega$ & 0 (fixed) & $-0.005_{-0.238}^{+0.212}$ & 0 (fixed) & $0.005_{-0.136}^{+0.131}$ & fitted \\ 
$\sqrt{e}\sin \omega$ & 0 (fixed) & $-0.472_{-0.119}^{+0.188}$ & 0 (fixed) & $-0.436_{-0.197}^{+0.446}$ & fitted \\ 
$T_{\rm{c}} $(BJD$-2450000$) & $4781.9_{-40.6}^{+22.9}$ & $4767.9_{-34.4}^{+45.1}$ & $4800.0_{-30.6}^{+29.5}$ & $4780.9_{-20.4}^{+26.2}$ & fitted \\ 
$e$ & 0 (fixed) & $0.223_{-0.072}^{+0.162}$ & 0 (fixed) & $0.190_{-0.134}^{+0.239}$ & derived \\ 
$\omega\ (^{\circ})$ & 0 (fixed) & $-90.6_{-28.3}^{+28.9}$ & 0 (fixed) & $-89.3_{-13.3}^{+99.3}$ & derived \\ 
$T_{\rm{p}} $(BJD$-2450000$) & $4660.4_{-36.5}^{+20.6}$ & $5009.2_{-472.0}^{+1.3}$ & $4679.9_{-31.7}^{+30.5}$ & $4540.6_{-3.0}^{+473.7}$ & derived \\ 
$M_{\rm{p}}\sin i\ (M_{\rm{J}})$ & $0.30_{-0.06}^{+0.03}$ & $0.32_{-0.05}^{+0.05}$ & $0.26_{-0.10}^{+0.08}$ & $0.28_{-0.10}^{+0.08}$ & derived \\ 
$a\ (\rm{au})$ & $1.334_{-0.005}^{+0.013}$ & $1.335_{-0.007}^{+0.006}$ & $1.327_{-0.009}^{+0.010}$ & $1.332_{-0.007}^{+0.006}$ & derived \\ 
\hline 
\textbf{Planet d} \\ 
$P\ (\rm{d})$  & $836.4_{-8.4}^{+8.4}$ & $838.0_{-7.0}^{+5.1}$ & $839.9_{-8.8}^{+10.1}$ & $837.7_{-5.5}^{+6.4}$ & fitted \\ 
$K\ (\rm{m\ s^{-1}})$ & $7.94_{-0.98}^{+0.74}$ & $8.62_{-0.97}^{+1.08}$ & $8.13_{-1.49}^{+1.45}$ & $8.73_{-1.69}^{+1.73}$ & fitted \\ 
$\sqrt{e}\cos \omega$ & 0 (fixed) & $0.061_{-0.178}^{+0.157}$ & 0 (fixed) & $0.006_{-0.134}^{+0.132}$ & fitted \\ 
$\sqrt{e}\sin \omega$ & 0 (fixed) & $-0.650_{-0.081}^{+0.151}$ & 0 (fixed) & $-0.499_{-0.206}^{+0.393}$ & fitted \\ 
$T_{\rm{c}} $(BJD$-2450000$) & $4926.0_{-23.8}^{+28.4}$ & $4911.2_{-64.8}^{+62.4}$ & $4924.5_{-25.3}^{+31.7}$ & $4925.8_{-26.5}^{+33.1}$ & fitted \\ 
$e$ & 0 (fixed) & $0.426_{-0.142}^{+0.132}$ & 0 (fixed) & $0.249_{-0.169}^{+0.259}$ & derived \\ 
$\omega\ (^{\circ})$ & 0 (fixed) & $-84.7_{-15.9}^{+14.4}$ & 0 (fixed) & $-89.3_{-42.4}^{+14.6}$ & derived \\ 
$T_{\rm{p}} $(BJD$-2450000$) & $4716.8_{-27.5}^{+30.1}$ & $4520.1_{-7.9}^{+815.5}$ & $4714.5_{-27.3}^{+33.4}$ & $4509.6_{-3.8}^{+822.3}$ & derived \\ 
$M_{\rm{p}}\sin i\ (M_{\rm{J}})$ & $0.45_{-0.06}^{+0.04}$ & $0.43_{-0.05}^{+0.05}$ & $0.46_{-0.08}^{+0.08}$ & $0.46_{-0.08}^{+0.08}$ & derived \\ 
$a\ (\rm{au})$ & $1.920_{-0.012}^{+0.012}$ & $1.923_{-0.011}^{+0.008}$ & $1.925_{-0.013}^{+0.015}$ & $1.922_{-0.008}^{+0.010}$ & derived \\ 
\hline 
\textbf{Extra parameters} \\ 
$s_{\rm{s}}\ (\rm{m\ s^{-1}})$ & $6.33_{-0.45}^{+2.14}$ & $6.72_{-1.17}^{+1.43}$ & $1.25_{-0.22}^{+1.43}$ & $1.28_{-0.25}^{+1.62}$ & fitted \\ 
$s_{\rm{f1}}\ (\rm{m\ s^{-1}})$ & $5.53_{-0.44}^{+1.08}$ & $5.45_{-0.73}^{+0.77}$ & $1.16_{-0.13}^{+0.86}$ & $1.19_{-0.16}^{+0.97}$ & fitted \\ 
$s_{\rm{f2}}\ (\rm{m\ s^{-1}})$ & $3.34_{-1.39}^{+1.40}$ & $1.72_{-0.67}^{+1.60}$ & $1.12_{-0.10}^{+0.61}$ & $1.14_{-0.12}^{+0.76}$ & fitted \\ 
$\gamma_{\rm{s}}\ (\rm{m\ s^{-1}})$ & $-1.69_{-1.23}^{+1.42}$ & $-1.16_{-1.30}^{+1.37}$ & $-1.97_{-1.62}^{+1.61}$ & $-1.82_{-1.70}^{+1.66}$ & fitted \\ 
$\gamma_{\rm{f1}}\ (\rm{m\ s^{-1}})$ & $-2.41_{-0.72}^{+0.94}$ & $-2.52_{-0.89}^{+0.86}$ & $-2.78_{-1.47}^{+1.49}$ & $-2.85_{-1.52}^{+1.52}$ & fitted \\ 
$\gamma_{\rm{f2}}\ (\rm{m\ s^{-1}})$ & $-1.06_{-0.75}^{+0.96}$ & $-0.94_{-0.90}^{+0.86}$ & $-1.17_{-2.07}^{+2.02}$ & $-1.18_{-1.96}^{+2.01}$ & fitted \\ 
\hline 
\textbf{Hyper parameters} \\ 
$\eta_{1}\ (\rm{m\ s^{-1}})$ & & & $6.60_{-0.82}^{+0.97}$ & $6.25_{-0.88}^{+1.05}$ & fitted \\ 
$\eta_{2}\ (\rm{d})$ & & & $30.0_{-10.6}^{+10.4}$ & $25.9_{-10.4}^{+13.3}$ & fitted \\ 
\hline\hline 
\end{tabular}}
\begin{tabnote}
\hangindent6pt\noindent
\hbox to6pt{\footnotemark[$*$]\hss}\unskip%
The symbol $s$ means extra jitter and the symbol $\gamma$ means RV offset to zero point. The lower subscript ``s'', ``f1'', and ``f2'' respectively refer to HIDES-S, -F1, and -F2.
\end{tabnote}
\label{tab:orbpar}
\end{table*}

To examine the fitting quality and perform model selection, we calculated root-mean-square (rms), reduced Chi-square $\chi^{2}_{\rm{red}}$ and Bayesian Information Criterion (BIC: \cite{Schwarz1978}) for each model, and we also performed dynamical stability analysis to rule out chaotic model (details in Section \ref{sec:dyn}).  
The reduced Chi-square is defined as 
\begin{eqnarray}
\chi^{2}_{\mathrm{red}}=\frac{\chi^{2}}{\nu},
\end{eqnarray}
where $\nu$ is the degree of freedom.
The uncertainty of $\chi^{2}_{\rm{red}}$ can be approximately given by $\sigma = \sqrt{2/n}$ \citep{Andrae2010}, where $n$ is the number of data. Therefore, we have $\sigma = \sqrt{2/180} = 0.105$ and a degeneracy within 3$\sigma$-interval of $0.684 \leq \chi^{2}_{\rm{red}} \leq 1.316$ for our HD 184010 data.
$\rm {BIC}$ is defined as 
\begin{eqnarray}
\mathrm{BIC} = k \ln{n} - 2\ln{\mathcal{L}},
\end{eqnarray}
where $\mathcal{L}$ is the maximized value of the likelihood function, $k$ is the number of parameters, and $n$ is the number of data.
A lower value of $\rm{BIC}$ indicates a more preferable model, therefore we calculate $\Delta\rm{BIC}$ of each model, defined by its own $\rm{BIC}$ minus the Null model's $\rm{BIC}$. 
A null model only fits RV offset $\gamma$ and extra jitter $s$ for each instrument.
A strong $\Delta\rm{BIC}$ evidence can be $-10 \leq \Delta\rm{BIC} \leq -6$ and a decisive $\Delta\rm{BIC}$ evidence can be $\Delta\rm{BIC} < -10$ \citep{Kass1996}.

\begin{table*}
\tbl{Model comparison}{%
\begin{tabular}{llccccr}
\hline\hline
Model & Planets & $\rm{rms}\ (\rm{m\ s^{-1}})$ & $\chi^{2}_{\rm{red}}$ & $\rm{BIC}$ & $\Delta \rm{BIC}$ & Stability\\ 
\hline
Null & No planet & $12.12$ & $1.039$ & $1439$ & $0$ & \\
$(e=0)$ & Planet b & $9.87$ & $1.056$ & $1394$ & $-45$ &\\
$e$ free & Planet b & $9.86$ & $1.084$ & $1394$ & $-45$ &\\
$(e=0)$ & Planet b and d & $8.10$ & $1.074$ & $1352$ & $-58$ & \\
$e$ free  & Planet b and d & $7.74$ & $1.107$ & $1335$ & $-104$ & \\
\textbf{(\textit{e} = 0)} & \textbf{Planet b, c, and d} & \textbf{7.01} & \textbf{1.112} & \textbf{1326} & \textbf{-113} & \textbf{Stable}\\
$e$ free  & Planet b, c, and d & $6.68$ & $1.103$ & $1306$ & $-133$ & Chaotic\\
$(e=0)$ + GP & Planet b, c, and d  & $3.86$ & $0.639$ & $1314$ & $-125$ & Stable\\
$e$ free + GP & Planet b, c, and d & $3.77$ & $0.636$ & $1305$ & $-134$ & Chaotic\\
\hline
\end{tabular}}
\begin{tabnote}
\hangindent6pt\noindent
\hbox to6pt{\footnotemark[$*$]\hss}\unskip%
The models are listed from the simplest one to the most complex one in terms of number of planets. \\
The signal of planet d ($P_{\rm{d}} \sim 840\ \rm{d}$) is strong than c ($P_{\rm{c}} \sim 480\ \rm{d}$), therefore we include planet d rather than c in the 2-Keplerian model. 
\end{tabnote}
\label{tab:modcomp}
\end{table*}

We list our model comparison and their results in Table \ref{tab:modcomp}.
The $\chi^{2}_{\rm{red}}$ values of all models without GP are approximately equal to 1, suggesting that neither under-fitting nor over-fitting occurred to these models.
The $\rm{BIC}$ values of planet models are all significantly lower than the Null model over a decisive level. 
Comparing planet models (1-, 2- and 3-Keplerian), adding planets reduces $rms$ and decreases $\rm{BIC}$ values significantly.
Thus 3-Keplerian models are much preferred to fit the regular RV variations in the RV time series.

Among four different 3-Keplerian models, both GP models have $\chi^{2}_{red} \sim 0.64$, which is out of the 3$\sigma$-interval of a good fit. 
This significance implies either the GP kernel improperly fits noise, or the error variance is overestimated. 
Thus we conclude that red noise cannot be characterized by the observed data, and a simpler model without GP should be selected.
Comparing $e=0$ Model and $e$ free model, $\rm{BIC}$ significantly favors the $e$ free Model.
Nonetheless, dynamical stability analysis (Section \ref{sec:dyn}) can convincingly rule out the $e$ free Model due to its chaos, while it can preserve the $e=0$ Model with its long-term stability.
Furthermore, we figure out that extra jitters, i.e., stellar jitter and instrumental jitter, can result in large eccentricity in orbit fitting. 
Using simulations, we can reproduce large eccentricity solutions by properly injecting noises to a $(e=0)$ Model, and provide details of our simulations are given in Appendix \ref{sec:sim}.

To sum up, the best-fit $e=0$ Model (Figure \ref{fig:rv}) is more preferable, having periods of $P_{\rm{b}}=286.6_{-0.7}^{+2.4}$ days, $P_{\rm{c}}=484.3_{-3.5}^{+5.5}$ days, and $P_{\rm{d}}=836.4_{-8.4}^{+8.4}$ days, semi-amplitudes of $K_{\rm{b}}=7.70_{-0.96}^{+0.78}\ \rm{m\ s^{-1}}$, $K_{\rm{c}}=6.32_{-1.06}^{+0.81}\ \rm{m\ s^{-1}}$, and $K_{\rm{d}}=7.94_{-0.98}^{+0.74}\ \rm{m\ s^{-1}}$ for planet b, c, and d, respectively. 
These values refer to semimajor axes of $a_{\rm{b}} = 0.940_{-0.001}^{+0.005}$ au, $a_{\rm{c}} = 1.334_{-0.005}^{+0.013}$ au, and $a_{\rm{d}} = 1.920_{-0.012}^{+0.012}$ au.
By using a stellar mass of $1.35M_{\odot}$, the minimum masses of three planets are $M_{\rm{b}}\sin i = 0.31_{-0.04}^{+0.03} M_{\rm{J}}$, $M_{\rm{c}}\sin i = 0.30_{-0.06}^{+0.03} M_{\rm{J}}$, and $M_{\rm{d}}\sin i = 0.45_{-0.06}^{+0.04} M_{\rm{J}}$.

The $e=0$ Model has rms scatter of $7.01\ \rm{m\ s^{-1}}$ for its residuals, and it has fitted extra jitters of $s_{\rm{s}} = 6.33_{-0.45}^{+2.14} \rm{m\ s^{-1}}$, $s_{\rm{f1}} = 5.53_{-0.44}^{+1.08} \rm{m\ s^{-1}}$, and $s_{\rm{f2}} = 3.34_{-1.39}^{+1.40} \rm{m\ s^{-1}}$ for HIDES-S, -F1, and -F2, respectively.
These results could be legitimately boiled down to solar-like oscillation and instrumental jitter. 
The expected RV jitter of solar like oscillation for HD 184010 is $A_{v} \sim 2.3\ \rm{m\ s^{-1}}$ \citep{Kjeldsen1995} or $A_{v} \sim 3.7 \ \rm{m\ s^{-1}}$ \citep{Kjeldsen2011}. 
The rms scatter for HIDES is approximately $\sigma \lesssim 4.5 \rm{m\ s^{-1}}$ by using RV standard star $\tau$ Cet \citep{Teng2022}.

\section{Line profile and chromospheric activity}\label{sec:line}
\begin{figure*}
\begin{center}
\includegraphics[scale=0.4]{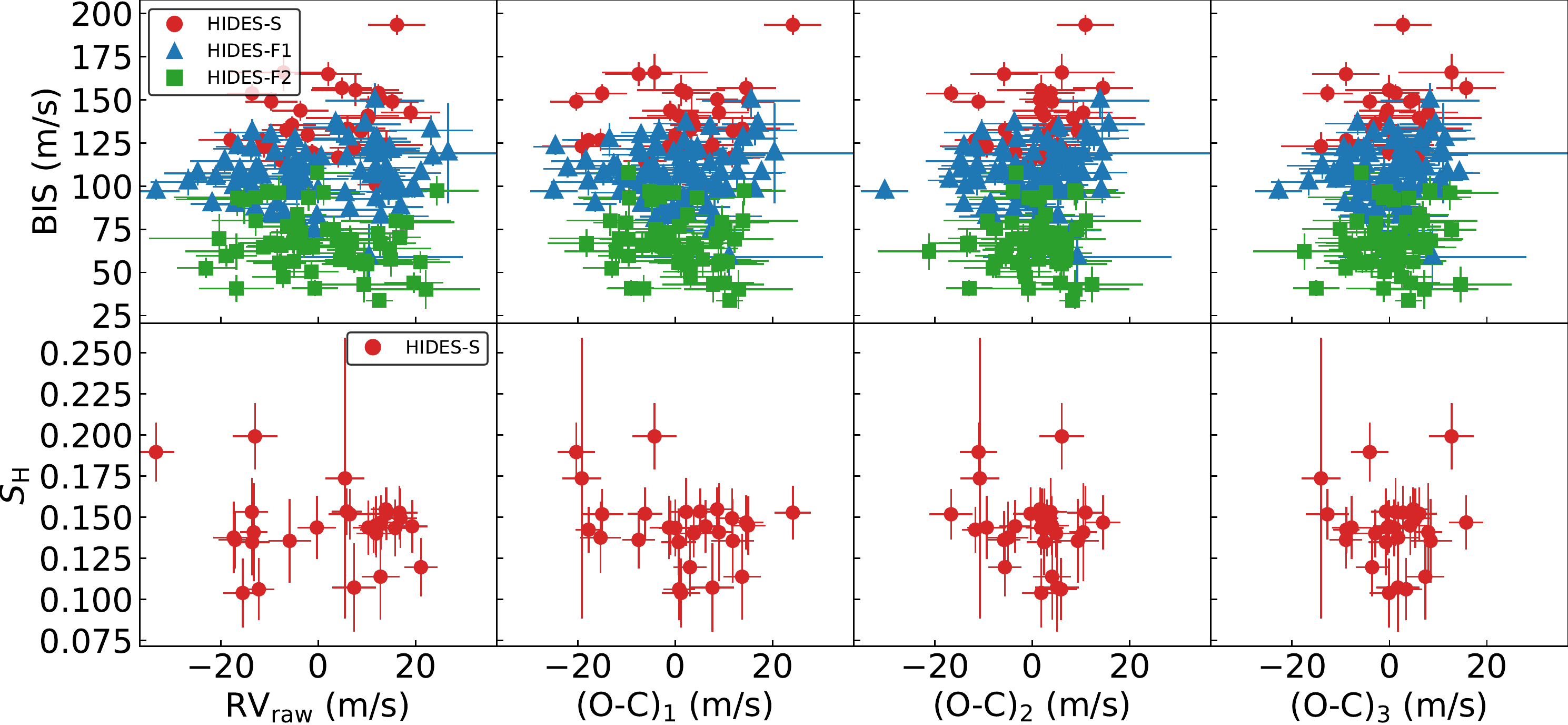} 
\end{center}
\caption{
BIS and index of Ca \emissiontype{II} H line $S_{\rm{H}}$ respectively against the observed RVs (RV$_{\rm{raw}}$), the residuals to 1-, 2-, and 3-Keplerian fittings [(O-C)$_{\rm{1}}$, (O-C)$_{\rm{2}}$, and (O-C)$_{\rm{3}}$] based on $e=0$ Model.
In all subplots, the symbols are the same as those in Figure \ref{fig:rv}. 
(A colored version of this figure is available in the online journal.)
}\label{fig:line}
\end{figure*}
Spectral line profile deformation can lead to wavelength shift and mimic planetary signals.
Therefore we performed line profile analyses in order to eliminate any fake planetary signal. We adopted bisector inverse span (BIS: \cite{Dall2006}) as an indicator of line profile asymmetry following the same way in \citet{Takarada2018,Teng2022}. In this measurement, the spectra were taken within the wavelength range of 4000-5000 \AA, in other words, the iodine-free region. We first calculated the weighted cross-correlation function (CCF: \cite{Baranne1996,Pepe2002}) constructed by shifting the mask as a function of Doppler velocity.
Then, we generated a numerical mask from \texttt{SPECTRUM} \citep{Gray1994} for a G-type giant star with approximately 800 lines. 
Finally, we calculated the BIS of the CCF. The BIS is defined as the offset of averaged velocity between the upper region (5\%--15\% from the continuum of the CCF) and the lower region (85\%--95\% from the continuum of the CCF).
Moreover, the difference in instrumental profile could lead to BIS offsets between instruments. Thus we defined mean removed BIS $\rm{BIS}^{\prime}=\rm{BIS}-\overline{BIS}$, where overline represents the mean value, in order to suppress the difference between instruments.

Stellar chromospheric activity can also result in RV variations. 
The flux of line cores of Ca \emissiontype{II} HK lines is an efficient indicator in the visible band to trace chromospheric activity \citep{Duncan1991}.
We use the Ca \emissiontype{II} H index $S_{\rm{H}} = F_{\rm{H}}/({F_{\rm{B}}+F_{\rm{R}}})$, where $F_{\rm{H}}$ is a total flux in a 0.66 \AA wavelength ~wide bin centered on the H line, $F_{\rm{B}}$ and $F_{\rm{R}}$ are respectively those in 1.1 \AA ~wide bins centered on minus and plus 1.2 \AA ~from the center of the H line \citep{Sato2013a}. The errors were estimated based on photon noise. Ca \emissiontype{II} K lines are not included due to their low signal-to-noise ratio in the HIDES-S spectra, We notice to readers that the order overlapping regions in HIDES-F1 and -F2 spectra (mentioned in Section \ref{sec:orbfit}) include Ca \emissiontype{II} HK lines, therefore we cannot obtain $S_{\rm{H}}$ from HIDES-F1 and -F2 spectra.

For HD 184010, we calculated Pearson’s correlation coefficient $r$ of BIS of each instrument against each set of RVs. Here RVs included observed raw RVs and RV residuals of 1-, 2- and 3-Keplerian $(e=0)$ Models. As a result, we obtain a weak correlation of $-0.3\lesssim r \lesssim 0.3$ for all data sets. We also noticed the large BIS offsets between HIDES-S, -F1, and -F2 (Figure \ref{fig:line}). Therefore we could know that the line shape asymmetry of this star was dominated by instrumental profile. 
We performed period analysis to BIS' and $S_{\rm{H}}$ yet the GLS periodogram showed no significant regular variation relevant to the three prospective planetary signals (Figure \ref{fig:rvls}). 
The maximum rotational period can be estimated from the projected rotational velocity of star $v \sin i$ and the star's radius. Using the values given in Table \ref{tab:properties}, we obtained a rotational period of $P_{\rm{rot}}\lesssim 184$ days, which is apparently shorter than the orbital period of 286.6 days.
A spotted star's rotational period could be deduced from photometry.
Therefore we also calculated GLS periodograms of \textit{Hipparcos}, ASAS-3, and ASAS-SN photometry and found no significant signal in them (Figure \ref{fig:rvls}).
Combining these facts, we conclude the three regular variations ($P=$ 286.6, 484.3, and 836.4 days) from the RV time series are best explained by the orbital-motion hypothesis, rather than a distortion of the line profile of stellar activity.

\section{Dynamical analysis}\label{sec:dyn}
In order to further constrain orbital parameters following Section \ref{sec:orbfit} and investigate the orbital stability of the system, we here present a dynamical study of the HD 184010 system following the same method presented in \citet{Sato2013b}, \citet{Sato2016}, and \citet{Takarada2018}. 
We initialize the system from the best-fit orbital status of 3-Keplerian $(e=0)$ Model derived in Section \ref{sec:orbfit} except for eccentricity since they were fixed to zeroes. 
Here we assume that all planets are co-planar and prograde, and integrate the system with a fourth-order Hermite scheme \citep{Kokubo1998}. 

\begin{figure}
\begin{center}
\includegraphics[scale=1]{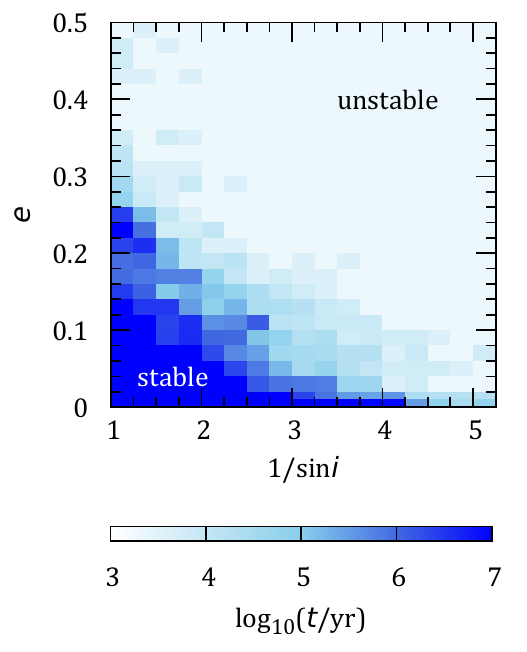} 
\end{center}
\caption{Lifetime of the system represented by eccentricity $e$ and multiple of minimum mass of planet $1/ \sin i$, where $i$ is the inclination angle. In the integration, planets are assumed to be coplanar and prograde, and they have longitude of pericenters initialized from zero.
The map gives the time elapsing before the semimajor axis of one planet deviates by 10\% from its initial value. 
The system is numerically integrated within 10 Myr. 
Thus the lifetime of 10 Myr here means that the system remains stable over 10 Myr years (dark-blue regions, marked by ``stable'') and lifetime of 1 kyr here means the system fails in keeping stable within 1 kyr (light-blue regions, marked by ``unstable''). 
In this lifetime map, each grid is set to be $e=0.02$ multiplying $1/ \sin i=0.2$ when $e > 0.02$, and $e=0.01$ multiplying $1/ \sin i=0.2$ when $e\leq 0.02$, respectively.
(A colored version of this figure is available in the online journal.)}\label{fig:stability_map}
\end{figure}

Figure \ref{fig:stability_map} is the stability map of the system, illustrating the lifetime of the system, and calculated by the 10-Myr integration. The lifetime here is defined as the time elapsing before the semimajor axis of one planet deviates by 10\% from its initial value.
We free the initial orbital parameters which are shown in each axis in the figure \ref{fig:stability_map}, i.e. eccentricity $e$ and the multiple of the minimum mass of planet $1/\sin i$ with longitude of pericenter was initialized from zero. Consequently, we obtain the lifetime map for HD 184010. 
As shown in Figure \ref{fig:stability_map}, the ``stable'' initials locate on the lower-left corner of the map, in other words, low mass and low eccentricity region, 
and ``stable'' initials include the best-fit three planet $e=0$ Model derived in Section \ref{sec:orbfit}.
This area can be approximately described by $(1/\sin i) + 20 e -2.8 < 0$ when initial eccentricity $e \ge 0.01$. 
Therefore, we know that the system may keep stable with relatively large initial eccentricities ($e\gtsim0.1$) as long as their masses are low enough, i.e., they may have weak mutual perturbations. 
From the figure, we also know that, if we initialize the system with very low eccentricities $e \leq 0.01$, the system can maintain stable even if the $1/\sin i$ values are as large as 4.2. That is to say, the true masses are 4.2 times to their minimum, with their inclinations $i \sim 14^{\circ}$. 
In addition, we also performed an examination in resonance, yet it returns neither two pairs nor the trio are in MMR.

\begin{figure*}
\begin{center}
\includegraphics[scale=0.34]{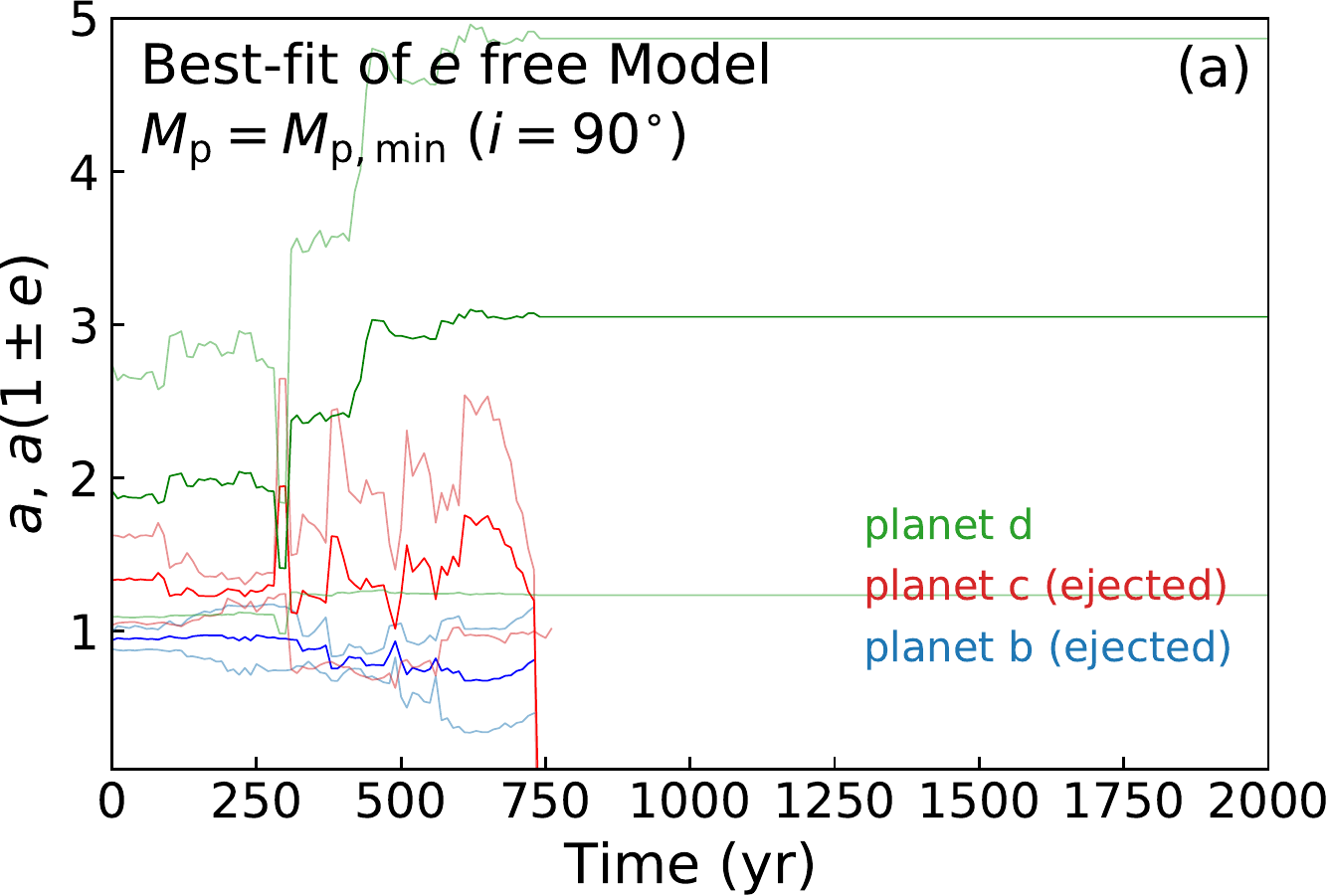}
\includegraphics[scale=0.34]{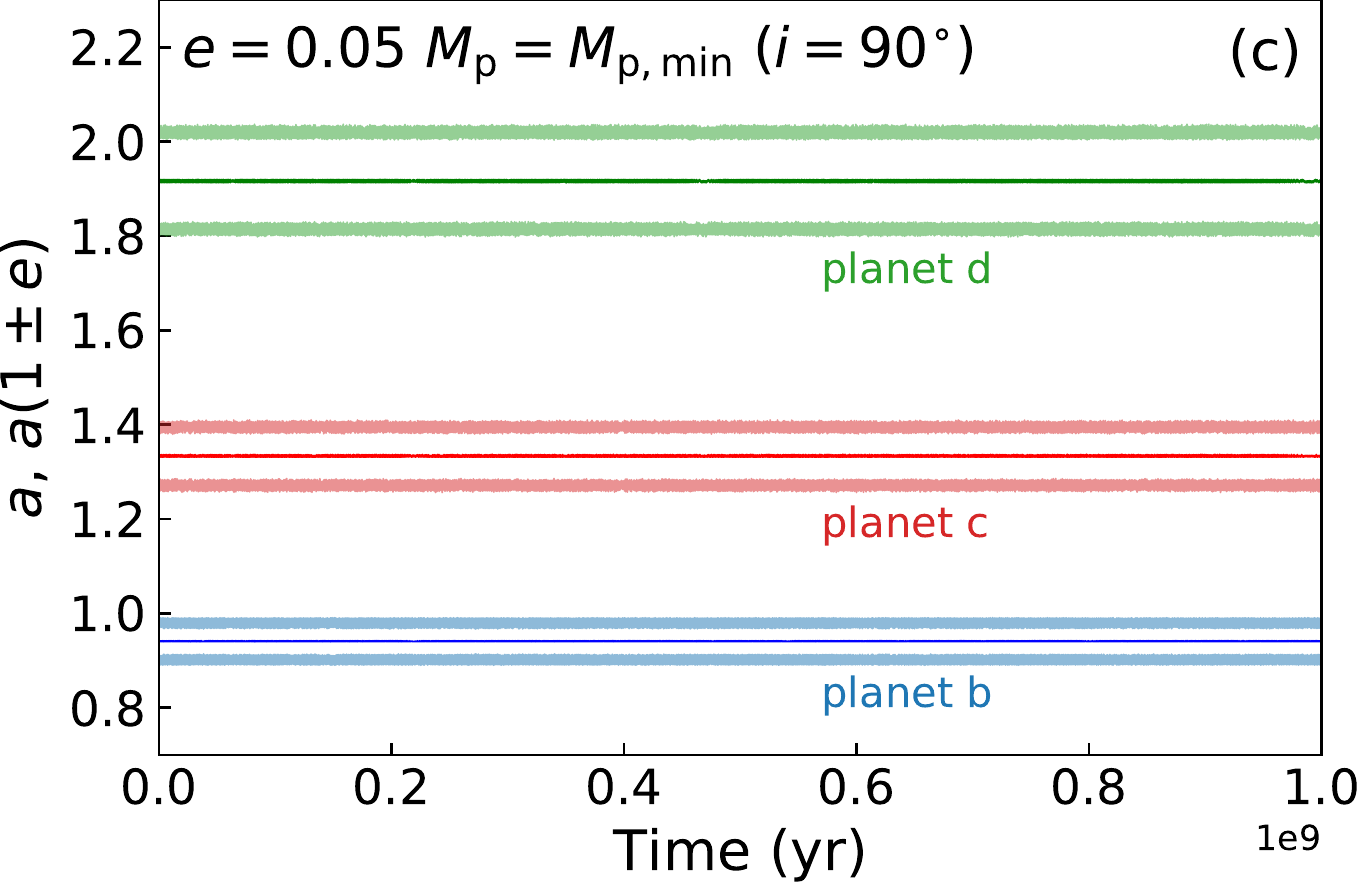}
\includegraphics[scale=0.34]{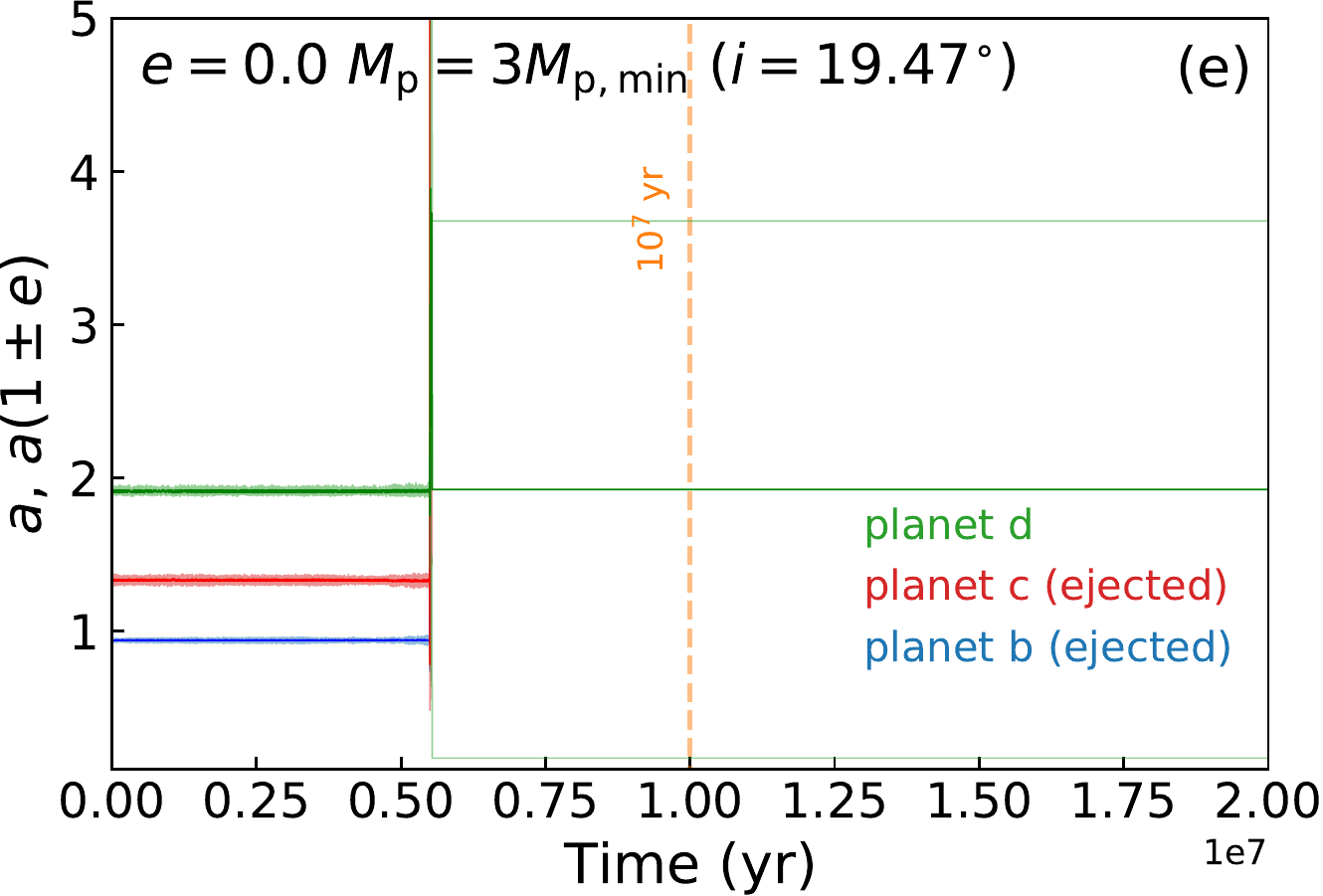}
\includegraphics[scale=0.34]{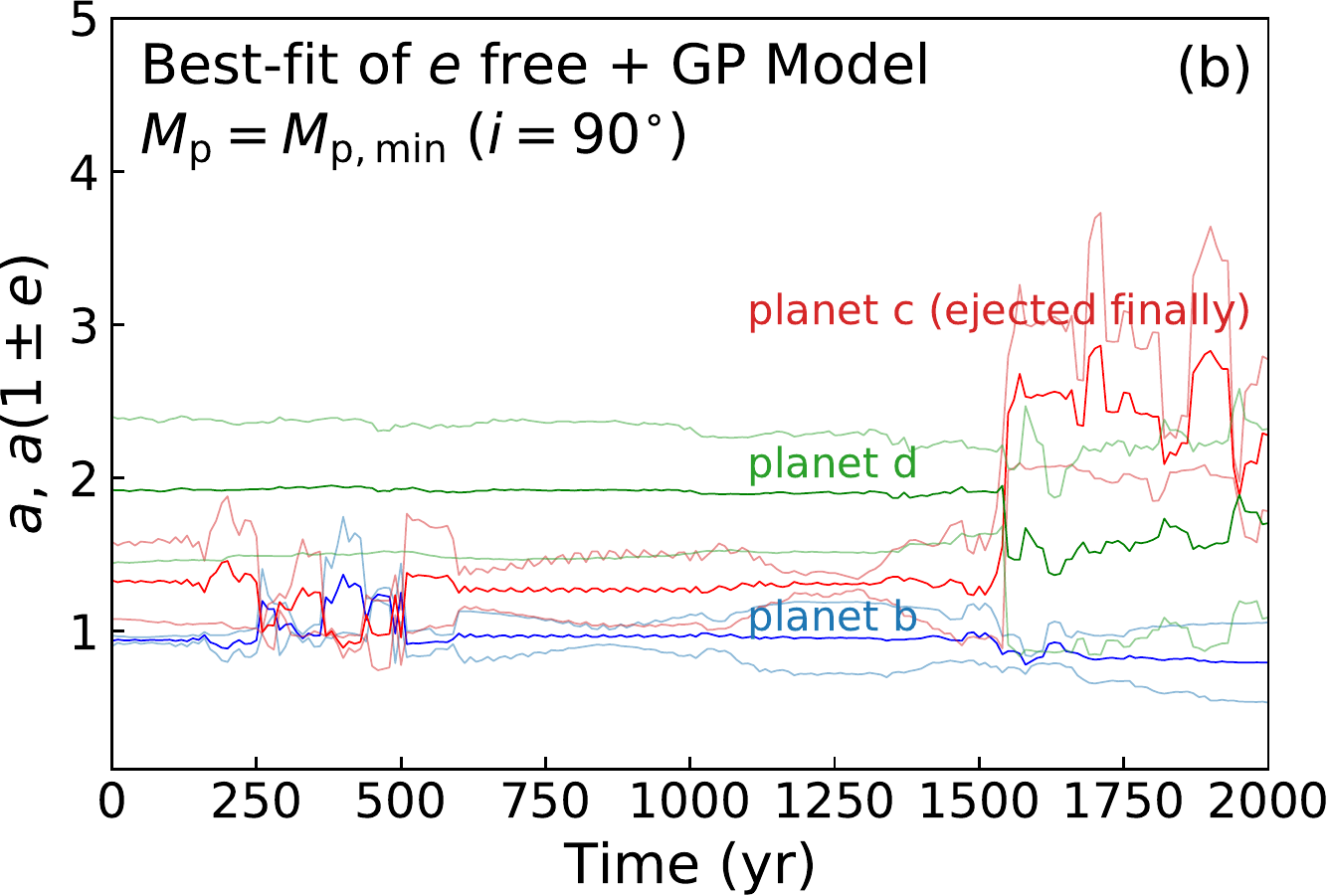}
\includegraphics[scale=0.34]{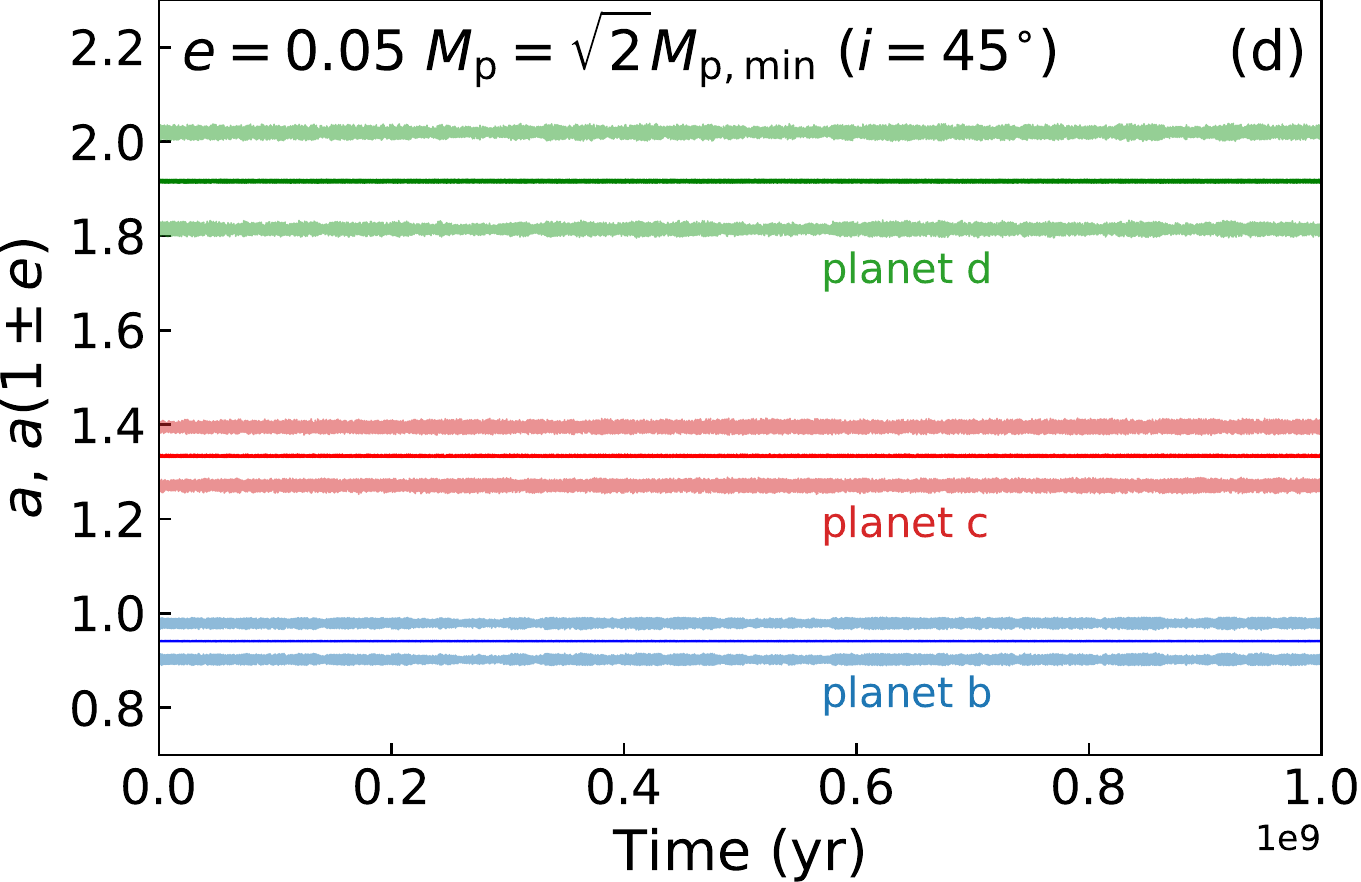}
\includegraphics[scale=0.34]{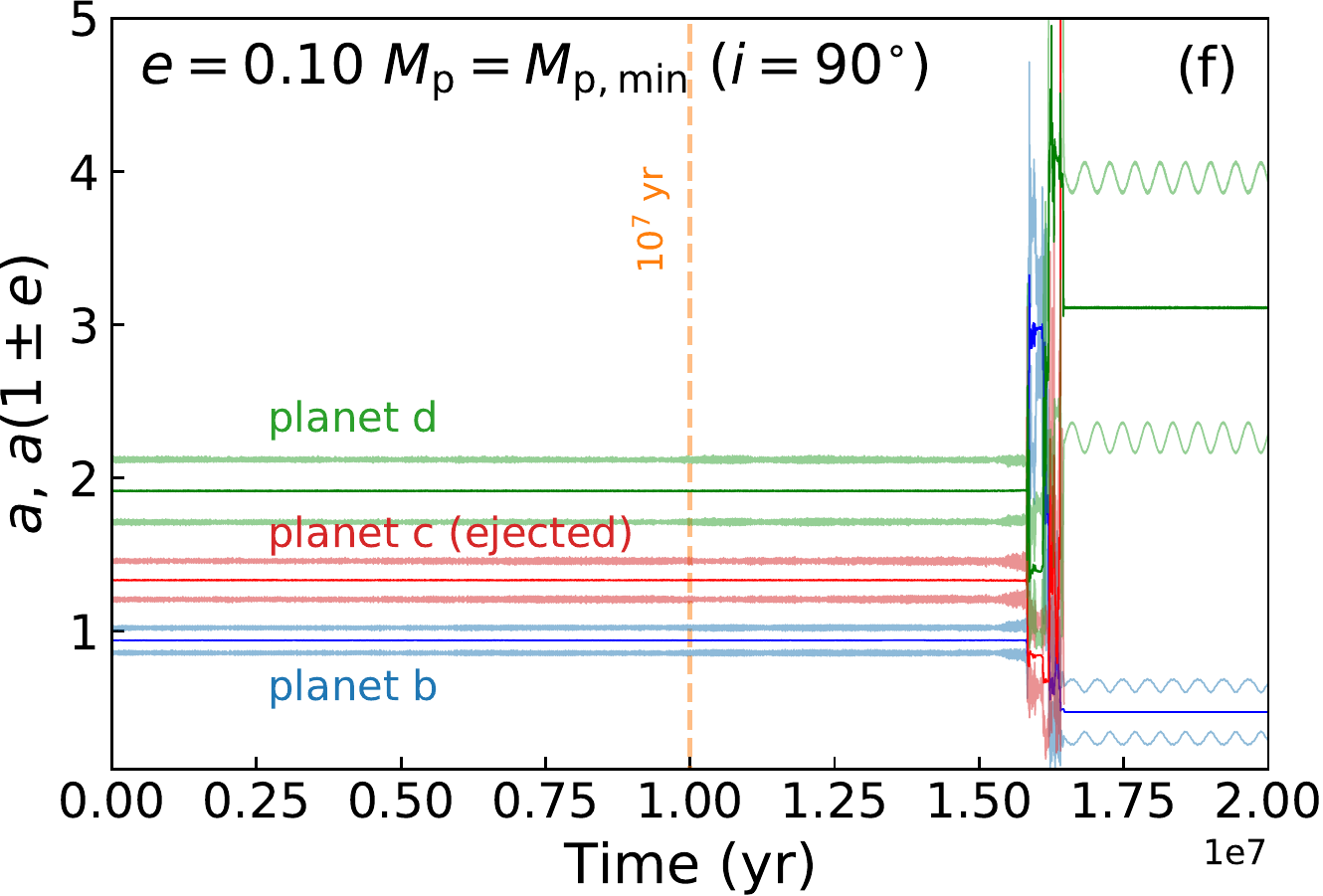}
\end{center}
\caption{
Evolution of semimajor axis $a$, pericenter distance $a(1-e)$, and apocenter distance $a(1+e)$ of the three planets in HD 184010 system. 
For subplot (a, b), they show the simulation initialized from best-fit $e$ free Model and $e$ free $+$ GP Model. Both of the orbits were unstable and the system quickly dissipated.
For subplot (c, d, e, f), they have: 
1) semimajor-axes $a$, minimum planet masses $M_{\rm{p, min}}=M_{\rm{p}}\sin i\ (i=90^{\circ})$ set to the best-fit 3-Keplerian $(e=0)$ Model; 
2) equal initial eccentricities; 
3) co-planar and prograde condition with random longitude of pericenters;
4) integration time of 1 Gyr.
The simulations in subplot (c, d) did not return any close encounter before the integration ended, while the simulations in subplot (e, f) returned close encounters with planet(s) ejected from the system. We only display the first 20-Myr evolution of orbital parameters in subplot (e, f) for a better readability.  
In all subplots, lines in darker colors are semimajor axes, and lines in lighter ones are pericenters (lower line for each planet) and apocenters (upper line for each planet) distance. 
(A colored version of this figure is available in the online journal.)}\label{fig:orb_evolution}
\end{figure*}

The stability map returned us a wide range of ``stable'' solutions, 
however, it assumed that planets had same initial eccentricities and the integration timescale was much shorter than the age of HD 184010 ($2.76$ Gyr: Section \ref{sec:star}).
We thus performed individual N-body simulations initialized from the best-fit $e$ free Model and $e$ free + GP Model derived in Section \ref{sec:orbfit} and several independent N-body simulations with ``stable'' solutions derived from the stability map with much longer time scales (1 Gyr) with could be more comparable to the age.
We applied \texttt{MERCURY6} code \citet{Chambers1999} and integrated the system with Bulirsch-Stoer method \citep{Chambers1998}. Here, we show six typical simulations with evolution of semimajoraxes $a$, pericenters $a(1-e)$, and apocenters $a(1+e)$ in Figure \ref{fig:orb_evolution}. 

The first two simulations (subplot a and b) were initialized from the best-fit $e$ free Model and $e$ free + GP Model, but neither of them could keep stable over the a time scale of $\sim 1$ kyr.
For $e$ free Model initials, planet ejection happened immediately after 700 yr, and for $e$ free + GP Model initials, one planet was pushed into a wider orbit after 1500 yr and finally ejected from the system.
These dynamical simulation could convincingly rule out the best-fit $e$ free Model and best-fit $e$ free $+$ GP Model although they had better fits under a BIC selection.

The latter four simulations (subplot c, d, e, and f) adopted the semimajor axes $a$ and minimum planet masses $M_{\rm{p, min}}=M_{\rm{p}}\sin i\ (i=90^{\circ})$ of the best-fit 3-Keplerian $(e=0)$ Model in Section \ref{sec:orbfit}. 
The initial eccentricities are equally set to all planets in each simulation, the initial of pericenters for eccentric orbits are set to be random values, and the system is set to be co-planar and prograde. 

For the simulation initialized with $e=0.05$ and $M_{\rm{p}}= M_{\rm{p, min}}$ ($i=90^{\circ}$) and $M_{\rm{p}}=\sqrt{2}M_{\rm{p, min}}$ ($i=45^{\circ}$), the system kept stable for over 1-billion year, with semimajor axes $a$, pericenters $a(1-e)$, and apocenters $a(1+e)$ slightly oscillating around their initial values.
But when the mass of the planets was increased to 3 times to their minimal ($i=19.47^{\circ}$), even in zero eccentricity orbits, the stability could be destroyed at around 5 Myr with close encounter happening. The inner two planets were ejected out of the system, and the outer one remained in system in an highly eccentric orbit.
This implied the true masses should be lower than 3 times to the minima. Therefore, we could obtain a tighter constraint on ranges of the three planets: $ 0.31 \leq M_{\rm{b}} \leq 0.93 $, $0.30 \leq M_{\rm{c}} \leq 0.90$, and $0.45 \leq M_{\rm{d}} \leq 1.35$. 
For the simulation initialized with $e=0.1$ and $M_{\rm{p}}= M_{\rm{p, min}}$ ($i=90^{\circ}$), the system kept stable within the first 15 Myr, a time longer than the integration timescale of stability map (Figure \ref{fig:stability_map}). But the strong interactions still triggered instability, forcing the planet in the middle ejected out of the system, with the other two planets pushed into new stable orbits with larger eccentricities thereafter.

These long-time-scale independent simulations suggest the parameter range of ``stable'' orbits would be smaller than what we obtained from the stability map. If the system have survived for a timescale of 1 Gyr, the orbits should be at least less eccentric with $e<0.1$, and the masses of planets should be more close to their minima, i.e. the system should be more edge-on than face-on to us.


\section{Discussion and summary}\label{sec:discuss}
In this paper, we report a trio of giant planets orbiting a K0 evolved star HD 184010 ($M_{\star} = 1.35 M_{\odot}$, $\log g_{\star} = 3.18$) by precise RV measurement from Okayama Planet Search Program. 
We used different models to explain the observed RV data, and consequently, we determined that the three planets should reside in near-circular orbits, with orbital periods of $P_{\rm{b}}=286.6_{-0.7}^{+2.4}\ \rm{d}$, $P_{\rm{c}}=484.3_{-3.5}^{+5.5}\ \rm{d}$, and $P_{\rm{d}}=836.4_{-8.4}^{+8.4}\ \rm{d}$, and semimajor axes of  $a_{\rm{b}} = 0.940_{-0.001}^{+0.005}$ au, $a_{\rm{c}} = 1.334_{-0.005}^{+0.013}$ au, and $1.920_{-0.012}^{+0.012}$ au, respectively.
Adopting the central star's mass, we derived planets' minimum masses of $M_{\rm{b}}\sin i = 0.31_{-0.04}^{+0.03} M_{\rm{J}}$ $M_{\rm{c}}\sin i = 0.30_{-0.06}^{+0.03} M_{\rm{J}}$, and $M_{\rm{d}}\sin i = 0.45_{-0.06}^{+0.04} M_{\rm{J}}$, respectively. 
We ruled out the orbital solutions with high eccentricities. Solutions with high eccentricities could not have long-term stability according to dynamical stability analyses, and the ``seeming'' high eccentricity could be mimicked by extra jitters.
HD 184010 system is unique and remarkable, and there is few resemblances from all the known planetary systems.
On the other hand, the orbital separations between the neighboring two planets are small. It is the only triple-giant-planet (or larger) system with all planets have intermediate periods ($10^2 < P < 10^3$ days). 
Thus it provides a fire-new example of the growing population of multi-giant-planet systems. 

\subsection{The uniqueness of HD 184010 system}
Among all previously known multi-planet systems around highly evolved stars (surface gravity $\log g < 3.5$, 25 systems in total), most of the systems are in the pattern of giant-planet pairs, and nearly half of them have small orbital separations with orbital period ratios close to or lower than 2:1 (Figure \ref{fig:logg_prot}). 

HD 33142 is the first highly evolved star discovered to have more than two non-transiting planets \citep{Bryan2016, Trifonov2022}. It is an intermediate-mass star at the very beginning of the red giant phase ($M_{\star}=1.52\ M_{\odot}$, $\log g = 3.375$ cgs). 
The system consists of three giant planets, including two Jupiter-like planets ($M_{\rm{b}} = 1.26\ M_{\rm{J}}$ and $M_{\rm{c}} = 0.89\ M_{\rm{J}}$) with orbital periods of $P_{\rm{b}}=330.0\ \rm{d}$ and $P_{\rm{c}}=810.0\ \rm{d}$ corresponding to semimajor axes of $a_{\rm{b}} = 1.074\ \rm{au}$ and $a_{\rm{b}} = 1.955\ \rm{au}$, respectively, and another inner Saturn-mass planet ($M_{\rm{d}} = 0.20\ M_{\rm{J}}$) with a period of $P_{\rm{b}}=89.9\ \rm{d}$ and an orbital separation of $a_{\rm{b}} = 0.452\ \rm{au}$. Compared with HD 184010, they have larger orbital separations between each other.

\textit{Kepler}-56 is another system having more than two planets \citep{Huber2013,Otor2016}. Although the host star has similar stellar properties ($M_{\star}=1.32\ M_{\odot}$, $\log g=3.31$ cgs) to HD 184010, the planetary system behaves quite different to HD 184010 does. A pair of transiting hot-Jupiters have masses of $M_{\rm{b}} = 0.070\ M_{\rm{J}}$ and $M_{\rm{c}} = 0.570\ M_{\rm{J}}$, and they orbit the host star at 0.103 and 0.165 au with period of 10.5 and 21.4 days respectively. The hot-Jupiter pair is misaligned with respect to the rotational axis of the host star. The further long-term RV follow-up reported a non-transiting gas giant (minimum mass of 5.61 $M_{\rm{J}}$) orbiting the host star in a much wider orbit at 2.16 au with a period of 1002 days. Although there is only one outer planet detected, \citet{Gratia2017} proposed that three outer planets are necessary for scattering to cause the amount of misalignment inferred for \textit{Kepler}-56’s planets b and c. 

Apart from these giant-planet-pairs, 7 CMa \citep{Wittenmyer2011,Luque2019} is the most similar one to HD 184010 in terms of stellar properties. The star is a K1 giant with mass $M_{\star}=1.34 M_{\odot}$ and surface gravity $\log g=3.19$, but more metal-rich ($\rm{[Fe/H]}=+0.21$) than HD 184010. The two planets are orbiting the host star closely in 4:3 ($<$ 2:1) resonance with periods of 980 and 745 days.

Considering main-sequence stars, planetary systems having three or more giant planets have been detected in different ways, and their host stars reveal a variety of stellar properties. Nonetheless, for those planetary hosts discovered by the RV method, HD 184010 is the second most massive one. Among trio systems, the host star $\upsilon$ And \citep{Butler1997,Butler1999} is a F8 dwarf having similar mass ($M_{\star} = 1.3 M_{\odot}$) to HD 184010. RV observations illustrates the star harbors one hot-Jupiter with minimum mass of $M_{\rm{b}}\sin i = 0.71 M_{\rm{J}}$ and orbital period of $P_{\rm{b}} = 4.6170$ days, one Jovian planet with minimum mass of $M_{\rm{c}}\sin i = 2.11 M_{\rm{J}}$ and orbital period of $P_{\rm{c}} = 241.2$ days, and another wide-orbit Jovian planet with minimum mass of $M_{\rm{d}}\sin i = 4.61 M_{\rm{J}}$ and orbital period of $P_{\rm{d}} = 1266.6$ days. Further astrometry observations measure inclinations and thus constrain the masses of $M_{\rm{c}} = 13.98_{-5.3}^{+2.3} M_{\rm{J}}$ and $M_{\rm{d}} = 10.25_{-3.3}^{+0.7} M_{\rm{J}}$. Other than HD 184010 system, three planets in $\upsilon$ And systems are in large separations, and a three-dimensional simulation investigated by \citet{Deitrick2015} provides the stable configurations of the system. According to robustly stable solutions, the hot-Jupiter should have a mass between 2 and 9 $M_{\rm{J}}$. 

More to the point, star $\upsilon$ And can be regarded as a progenitor, i.e. main-sequence star, of HD 184010. However, it would be hard to detect the HD 184010 system if we assume the HD 184010 is still at its main-sequence phase. For a typical young F-type dwarf, it intrinsically exhibits larger RV jitter than G- and K- type dwarfs \citep{Wright2005, Isaacson2010, Luhn2020}. Typically, the jitter could reach at least $5\ \rm{m\ s^{-1}}$ for an inactive star, while it could as large as 10 or several tens of $\rm{m\ s^{-1}}$ for a more active star. 
Considering the HD 184010 system, the planets can be only as massive as $\sim 0.4 M_{\rm{J}}$, and they excite RV semi-amplitude less than $8\ \rm{m\ s^{-1}}$. In such a case, RV detection of an HD 184010-like system around a dwarf star could probably be failed.
Thus, we might lose the chance to directly have a view of ``young'' HD 184010 systems, in other words, we might not directly see how the HD 184010-like systems evolve as the host stars evolve from RV surveys. 

As for other trio systems, HD 37124 (G4 dwarf, $M_{\star}=0.8 M_{\odot}$, \cite{Butler2003,Vogt2005}) system shares some similarities to HD 184010. The orbital periods of three planets in HD 37124 system ($P_{\rm{b,c,d}}=154,\ 885,\ 1862$ days) is comparable to the ones in HD 184010 systems, i.e. there is neither hot-Jupiters nor planets in wide orbits. For further systems, they have either wide orbit planets, e.g. 47 UMa \citep{Butler1996b,Fischer2002,Gregory2010}, or hot-Jupiters, e.g. the aforementioned \textit{Kepler}-56, HIP 14810 \citep{Butler2006,Wright2007,Wright2009}, HD 27894 \citep{Moutou2005, Trifonov2017}, and etc. 

\subsection{Formation and migration}
So far, it is unclear why multi-giant-planet systems, including giant-planet-pairs, with small separations and intermediate orbits, are preferentially detected around evolved stars more massive than the sun (Figure \ref{fig:logg_prot}).
In terms of star mass, it might be a primordial property that massive stars could result in this planet pattern since massive stars could provide more materials in planet formation. 
In terms of the star's evolution stage, on the one hand, it might be the observational bias among the existed surveys, because massive stars are not suitable targets for RV measurements at their main-sequence phase. That is, even if they exist, we can hardly find them. On the other hand, it also might be stellar tide and mass loss of the host star, which act on the evolution of the planet's orbit. 
Or, we may need to revisit planet formation/migration theories, which could be challenged by these systems.

In this study, it is difficult to draw a solid conclusion on the formation scenario of the HD 184010 system, or similar systems.
For the formation of giant planet systems, the majority would consider the core-accretion scenario, where the gas giant planets form via gas accretion around solid cores at the outer part of the disk beyond the ice boundary around 4 au \citep{Lissauer1995, Boss1995}.
However, gas in circumstellar disks dissipates rapidly after a few Myr, with remaining typically much less than the mass of Jupiter \citep{Zuckerman1995}. 
Therefore, it is questionable whether solid cores could accrete enough gas so that planets be formed on that timescale. 

HD 184010 has a trio of planets all residing inside 4 au, and the total mass of the planets is higher than a Jupiter mass. 
In other words, if core-accretion is the true scenario, the HD 184010 system provides evidence that main-sequence F-type stars, i.e. the progenitor of HD 184010, could provide sufficient materials for accretion procedure, and mass dissipation is not as fast as expected. 
Nonetheless, an inward migration procedure is necessary for this system.
Therefore, interaction with other objects is required to conserve the energy in the system, so that three known planets could move inwardly.

Type II migration could be a possible explanation for the inward migration of giant planets. 
The classical model follows the viscous disk evolution \citep{Lin1986} with a giant planet opening up a gap in the protoplanetary disk and migrating on the same time scale as the viscous time scale \citep{Ward1997}. More recent high-resolution hydrodynamical simulations show that the gap around Jupiter-mass planet orbit is not empty and the migration is not tied to the disk gas accretion (e.g. \cite{Kanagawa2015,Duffell2015,Kanagawa2018,Ida2018}).
Moreover, considering the planets in HD 184010 system are likely to be less massive than Jupiter, the system may undergo type III migration regime \citep{Masset2003}, which is driven by co-orbital torques and happen in a much shorter timescale.
This regime typically corresponds to Saturn-mass planets in massive protoplanetary disks, which seems to conform to the case of HD 184010 and its planets. Besides, for planets having a period ratio lower than 2:1, they should have experienced the special condition that triggers the planets passing through strong 2:1 resonance during the disk migration\footnote{Further investigation for these scenarios are out of the scope of this paper.}.

In addition, we also notice that the outermost planet in HD 184010 system is more massive than the inner two planets. It could be explained in this way. As planets are embedded in the disk, outer planets directly absorb gas from a larger ``gas reservoir'', i.e. outer region of the disk, and grow with a full disk accretion rate. However, the inner planets can only accrete gas that inwardly passed the outer ones with a lower gas accretion rate \citep{Lubow2006}.

Except for disk migration, the existence of a much more distant giant planet could be another plausible explanation for the inner migration of the three  detected planets. Yet, there is no observational evidence for a fourth planet in wide orbit from our 17-year RV time series. 
Probably, the large eccentricity of planet d from $e$ free Model might be caused by such an unseen planet, and this planet has a harmonic period with the known planet d. 
But at least, our analysis ruled out this scenario from the current RV data. 
Or, if we assume an undetected RV acceleration of $\dot{\gamma} = 0.1\ \rm{m\ s^{-1}}$ caused by Jupiter-mass planet, 
such a planet should reside over $a = 42\ \rm{au}$ away from the host star, corresponding period around 240 yr and a distance never suitable for RV measurements. 

Moreover, HD 184010 is a star that has just passed the subgiant branch and stepped into its red-giant-branch. Hence we here ignore its mass loss during its evolution to the current stage (See section \ref{sec:star}), that is to say, we ignore the outward migration of the three planets caused by the mass loss of the central star.

\subsection{Dynamical stability}
The stability of these multi-planet systems has been studied over years. 
\citet{Chambers1996} investigated the stability of systems of more than two planets using numerical integration and found systems with the semimajor-axis difference $\Delta < 10$ (measured by mutual Hill radius $R_{\rm{H}}$) are always unstable. Their simulations include a solar-mass host star and equally spaced $10^{-7} M_{\odot}$ planets on initially circular and co-planar orbits. According to the relationship between separation and instability timescale, the first encounter time may happen at around 100 Myr at $\Delta \sim 8$. 
\citet{Marzari2002} studied the situation of three $10^{-3} M_{\odot}$ planets equally spaced on initially circular and co-planar orbits, and they obtain a relation between separation and instability timescale for Jupiter-mass planets. 
They firstly studied the case of $3 \leq \Delta \leq 5.3$ with inner planet fixed at semimajor-axis $a = 5 \rm{au}$. At a maximum initial separation of $\Delta = 5.3$, the system can maintain stability for about 1 Gyr. 
Furthermore, assuming the \textit{in situ} formation of giant planets close to the host star from the disk, they initialized the planets respectively at semimajor-axis of 1, 1.5, and 2 au, and inclination of 0.5$^{\circ}$, 1.0$^{\circ}$, and 1.5$^{\circ}$, referring to $\Delta_{1,2} = 4.65$ and $\Delta_{2,3} < 3.32$ (the two subscript numbers represent planet numbers). The system became unstable after 30 kyr, which basically followed the prediction. 

Although HD 184010 system looks similar to this simulation assuming \textit{in situ} in \citet{Marzari2002}, the planet separations of HD 184010 are much larger if they are measured in mutual Hill radii. 
The separation of the inner two and outer two planets are respectively $\Delta_{1,2} = 6.56$ and $\Delta_{2,3} = 6.47$ for planets at their minimum masses (inclination $i= 90^{\circ}$), and $\Delta_{1,2} = 5.84$ and $\Delta_{2,3} = 5.76$ for planets at 1.4 times of their minimum masses (inclination $i= 45^{\circ}$). 
Our simulation in Section \ref{sec:dyn} convinced stability of at least 1 Gyr for the upper two orbital separations, reaching a consensus with the prediction by Figure 2 in \citet{Marzari2002}.
However, when the planets are increased to several times massive than their minima, e.g. three times (upper right subplot in Figure \ref{fig:orb_evolution}), the separation are decreased to $\Delta_{1,2} = 4.54$ and $\Delta_{2,3} = 4.48$. We can see that, the stable lifetime of around 5 Myr years also roughly follows the relation in order of magnitude.

Eventually, by considering the stellar age of around 3 Gyr, the HD 184010 system should survive the same time scale. In order to keep the system stable, the planets should have masses not far from the minima obtained from Keplerian fitting and the orbits should have small eccentricities. 


\begin{ack}
We thank the anonymous referee for the very valuable and important suggestions that allowed us to improve the paper.

This research is based on data collected at the Okayama Astrophysical Observatory (OAO), which was operated by the National Astronomical Observatory of Japan. 
We are grateful to all the staff members of OAO for their support during the observations.
The Okayama 188cm telescope is operated by a consortium led by Exoplanet Observation Research Center, Tokyo Institute of Technology (Tokyo Tech), under the framework of tripartite cooperation among Asakuchi-city, NAOJ, and Tokyo Tech from 2018.
We thank the students of Tokyo Institute of Technology and Kobe University for their kind help with the observations at OAO. 
We express our special thanks Yoichi Takeda for his support in stellar property analysis. 
B.S. was partially supported by MEXT's program ``Promotion of Environmental Improvement for Independence of Young Researchers" under the Special Coordination Funds for Promoting Science and Technology, and by Grant-in-Aid for Young Scientists (B) 17740106 and 20740101, Grant-in-Aid for Scientific Research (C) 23540263, Grant-in-Aid for Scientific Research on Innovative Areas 18H05442 from the Japan Society for the Promotion of Science (JSPS), and by Satellite Research in 2017-2020 from Astrobiology Center, NINS.
H.I. was supported by JSPS KAKENHI Grant Numbers JP16H02169, JP23244038.

In this research, we also adopted \texttt{NumPy} \citep{Harris2020} for polynomial fit and \texttt{SciPy} \citep{Virtanen2020} for likelihood maximization and MAP fitting. We make contour plots using \texttt{pytgc} \citep{Bocquet2016}.

This research has made use of the SIMBAD database, operated at CDS, Strasbourg, France. This work has made use of data from the European Space Agency (ESA) mission {\it Gaia} (\url{https://www.cosmos.esa.int/gaia}), processed by the {\it Gaia} Data Processing and Analysis Consortium (DPAC,
\url{https://www.cosmos.esa.int/web/gaia/dpac/consortium}). Funding for the DPAC has been provided by national institutions, in particular the institutions participating in the {\it Gaia} Multilateral Agreement.
This research has made use of the NASA Exoplanet Archive, which is operated by the California Institute of Technology, under contract with the National Aeronautics and Space Administration under the Exoplanet Exploration Program.

\end{ack}

\appendix

\section{$V$-band Photometry}\label{sec:photometry}
\begin{figure*}
\begin{center}
\includegraphics[scale=0.6]{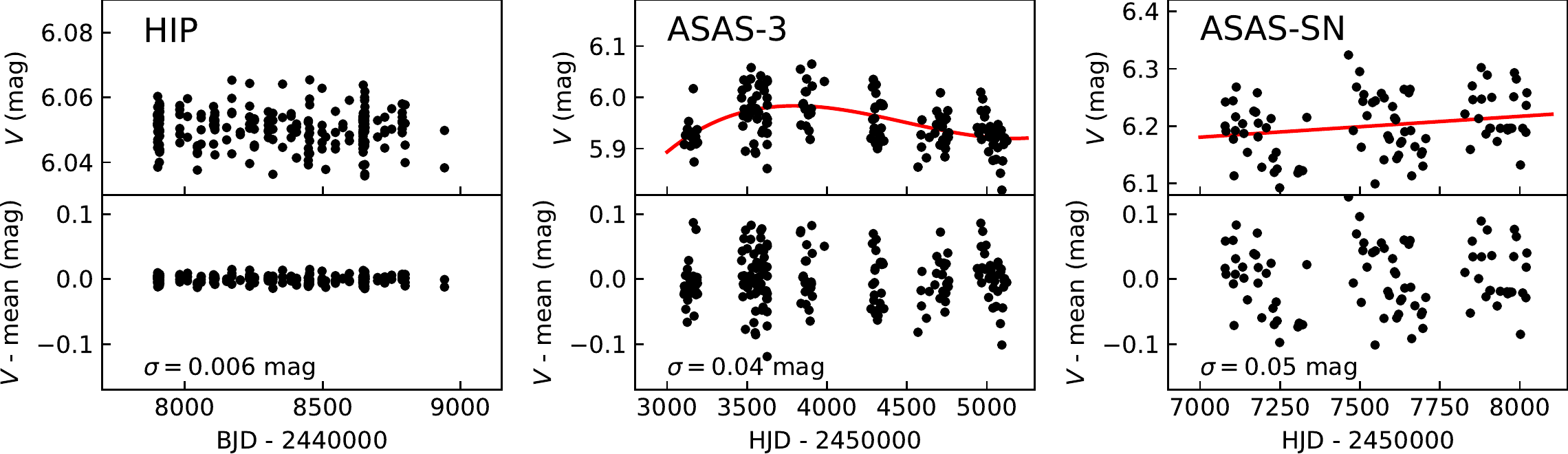} 
\end{center}
\caption{
$V$-band light curves obtained by \textit{Hipparcos} (left subplot), ASAS-3 (middle-subplot), and ASAS-SN (right subplot).
The upper subplot of each column is the non-detrended light curve, with its trend shown in a red solid curve.
The lower subplot of each column is the detrended light curve with its mean value removed.
(A colored version of this figure is available in the online journal.)}\label{fig:lc}
\end{figure*}
In this work, we adopted $V$-band photometry from three different surveys: \textit{Hipparcos}, ASAS-3, and ASAS-SN. For the ASAS-3 light curve, we adopted its photometry from its largest aperture. Although ASAS-SN officially warned  a probable saturation, we determined to adopt the data as a comparison.

The light curves were processed before calculating scatters and performing GLS periodograms. 
We first removed bad observations marked in the archival data, and we discarded the ASAS-3 light curve before HJD 2453000 due to the large scatter.
Since \textit{Hipparcos} has the highest long-term stability (a scatter of 0.02 mag before processing) among the three light curves, we decided to flatten the long-term trend in the other two light curves.
We used a 3rd order polynomial to fit the arched trend passing through the ASAS-3 light curve, which had a time span of 2016.6 d.
We removed the arched trend in the ASAS-3 light curve and a linear trend in the ASAS-SN light curve. 
Finally, we discarded data out of 3-$\sigma$ for all three light curves (Figure \ref{fig:lc}).

\section{Jitters mimic eccentric orbits}\label{sec:sim}
Jitters can mimic eccentric orbits when the observations are sparsely spaced and RV amplitude caused by orbital motion is relatively low. 

\subsection{HD 184010 case}
In the HD 184010 case, the best-fit $e=0$ Model has semi-amplitudes of $K_{\rm{b}}=7.70_{-0.96}^{+0.78}\ \rm{m\ s^{-1}}$, $K_{\rm{c}}=6.32_{-1.06}^{+0.81}\ \rm{m\ s^{-1}}$, and $K_{\rm{d}}=7.94_{-0.98}^{+0.74}\ \rm{m\ s^{-1}}$ for planet b, c, and d, respectively. 
These values are larger than, but comparable within errors to the observed data $\sim 4.5\ \rm{m\ s^{-1}}$, rms scatter of the 3-Keplerian residuals $\sim 7.01\ \rm{m\ s^{-1}}$, and fitted extra jitter $\sim 5\ \rm{m\ s^{-1}}$. 

Besides, the sampling of the observations is quite sparse. On average, it is one RV point per five weeks within the full time-span, and one RV point per three weeks after BJD 2456000 (early 2012).

Therefore, we decided to perform a series of simulations to investigate how jitters mimic eccentric orbits for HD 184010.

\subsection{Simulated RV data}
In order to keep a good consistency with the real HD 184010 RV data, we adopted the observational window and three instruments respect to real data for our simulations, and we empirically set the observational error to a random value $3.5 \sim 6\ \rm{m\ s^{-1}}$ based on the real errors. The simulated data were then generated as follows:

First, We first generated RV time series caused by orbital motions from the best-fit $e=0$ Model derived in Section \ref{sec:orbfit}. 
Then we injected different jitter components to the RV time series. 
We considered both white Gaussian noise and red Gaussian noise components, although red noise model was actually unknown for HIDES. 
The white Gaussian noises consisted of stellar jitter, which should be solar-like $p$-mode oscillation dominated\footnote{
The frequency of maximum power $\nu_{\rm{max}}$ of solar-like oscillation of HD 184010 should be approximately 180 $\rm{\mu Hz}$, which can be estimated from its stellar properties \citep{Kjeldsen1995}, referring to a period of 1.5 h, which is too short to be considered as a correlated noise
}, and instrumental jitter. 
The red Gaussian noises, probably due to instrumental drift and stellar activity, were purely hypothetical in our simulations. They were assumed to consist of four signals, having correlation function $e^{-|\tau|}$ with decorrelation time scales of 7 d, 30 d, 1 yr, and 3 yr, respectively. These jitters are added to data by each instrument. 
We tried two different sets of jitters, including a red-noise-dominated set (Model A) and a white-noise-dominated set (Model B). Also, we injected random RV offsets between instruments within $3\ \rm{m\ s^{-1}}$.
So far, a data generation process was finished.

We performed 3-Keplerian orbital fits with $e$ free Model (the same model in Section \ref{sec:sim}) to these simulated data. The best-fit of each simulation was determined by a Maximum A Posteriori (MAP) method using \textsf{Nelder-Mead} algorithm. In each fitting, we set an initial guess equal to the true orbital parameter set with a prior given in Table \ref{tab:prior}. Since error estimation and posterior distribution were not necessary for each simulation, we only derived the best fit and did not perform MCMC sampling for a time-saving reason. Furthermore, we ruled out fitting results including any best-fit eccentricity $e>0.8$.

Before extensively generating simulations, we first tried a few runs and calculated the rms of the residuals, and we slightly adjusted the scatters of jitter components, so that the rms of residuals from simulations could be at the same level as the rms value derived from real data. 
After we obtained a proper value for each jitter component, we generated 2000 simulations for each jitter model.
The adopted scatter of each jitter component is listed in \ref{tab:sim}

\subsection{Results and summary}
\begin{figure*}
\begin{center}
\includegraphics[scale=0.6]{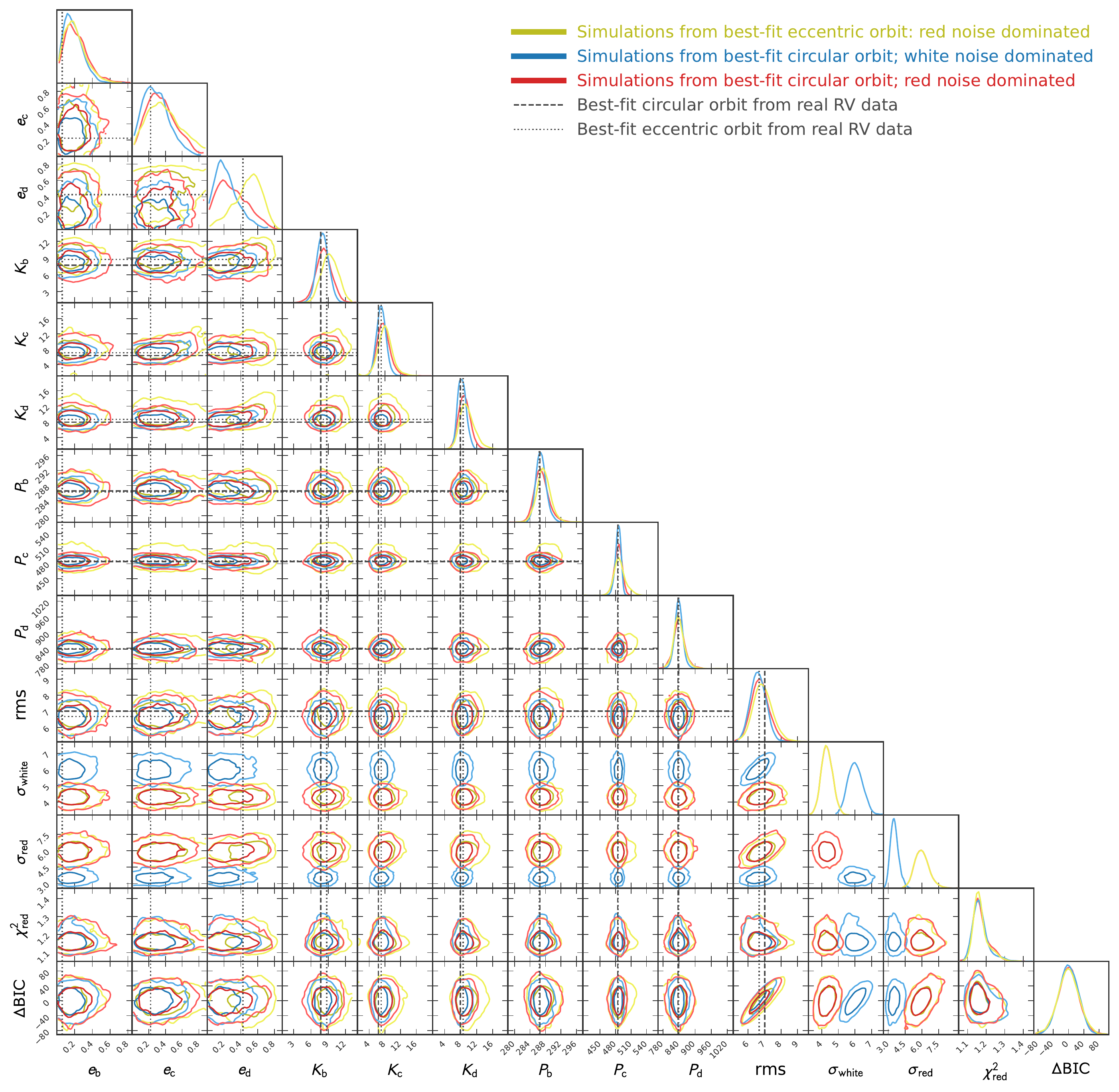} 
\end{center}
\caption{
Contour plots of fitting results from simulated RV data.
We display the distributions of best-fit orbital parameters from simulated RV data, including eccentricities ($e$), semi-amplitudes ($K$), and orbital periods ($P$).
We also display the distributions of other parameters from simulated RV data, including rms of the fittings, the scatter of simulated white noise ($\sigma_{\rm{white}}$), the scatter of simulated red noise ($\sigma_{\rm{red}}$), reduced Chi-square ($\chi_{\rm{red}}^{2}$), and the difference to the mean BIC value of all fittings. 
We use different colors to illustrate different sets of simulations. 
Red (Model A): circular orbit initialized and red noise dominated; 
Blue (Model B): circular orbit initialized and white noise dominated; 
Yellow (Model C): eccentric orbit initialized and red noise dominated. 
Dashed line is the best-fit circular orbit from real RV data, and dotted line is the best-fit eccentric orbit from real RV data.
(A colored version of this figure is available in the online journal.)}\label{fig:sim}
\end{figure*}

Consequently, we obtain a distribution of $\sim$2000 best fits for each jitter model in Figure \ref{fig:sim}, and specifically, we report the median and 1$\sigma$ for best-fit eccentricities $e_{\rm{b, c, d}}$ in Table \ref{tab:sim}.

The distribution for each model is almost Gaussian-like.
From the distribution, we find eccentricities $e$ largely biased from their initials, while the orbital periods $P$ and semi-amplitudes $K$ do not.
This especially happens for the outer two planets, whose eccentricities were significantly over-estimated by the true value 0 almost exceeding 2$\sigma$ of the distribution.
The bias happens to both Model A and B, suggesting that both red noise and white noise can mimic eccentricities. 
Red noise might affect eccentricity to a larger extent than white noise since Model A has larger biases of eccentricities compared with Model B.
Besides, we believe that these simulations on average have good fits with $\chi_{\rm{red}}^{2}$ at a level of 1.2.

In addition, we also performed simulations initialized by the best-fit $e$ free Model derived from real HD 184010 data in Section \ref{sec:orbfit}. The jitter model is the same as Model A, and other conditions are the same as Model A and B. 
As a result, we obtain even large eccentricities from highly eccentric initials, while orbital periods $P$ and semi-amplitudes $K$ do not largely bias from the initials (Table \ref{tab:sim} and Figure \ref{fig:sim}).

In summary, we conclude that jitters can mimic eccentric orbits, and we suggest that, for HD 184010, the large eccentricities from an $e$ free Model can be attributed to this scenario.

\begin{table*}
\tbl{Simulation parameters}{%
\begin{tabular}{lccc}
\hline\hline
Model & A & B & C \\
\hline
\textbf{Injected signals} \\
Orbital motion & best-fit $e=0$ Model & best-fit $e=0$ Model & best-fit $e$ free Model  \\
Main noise & Red & White & Red \\
$\sigma_{\rm{white, inst}} (\rm{m\ s^{-1}})$ & $3.0$ & $4.5$ & $3.0$ \\
$\sigma_{\rm{white, star}} (\rm{m\ s^{-1}})$ & $4.0$ & $5.5$ & $4.0$  \\
$\sigma_{\rm{red, 7d}} (\rm{m\ s^{-1}})$ & $2.5$ & $1.5$ & $2.5$ \\
$\sigma_{\rm{red, 30d}} (\rm{m\ s^{-1}})$ & $2.5$ & $1.7$ & $2.5$ \\
$\sigma_{\rm{red, 1yr}} (\rm{m\ s^{-1}})$ & $2.2$ & $1.2$ & $2.2$ \\
$\sigma_{\rm{red, 3yr}} (\rm{m\ s^{-1}})$ & $0.8$ & $0.3$ & $0.8$ \\
\hline
\textbf{Observed signals} \\
$e_{\rm{b}}$ (derived) & $0.20_{-0.11}^{+0.14}$ & $0.17_{-0.09}^{+0.11}$ & $0.19_{-0.10}^{+0.13}$ \\
$e_{\rm{c}}$ (derived) & $0.33_{-0.16}^{+0.20}$ & $0.27_{-0.14}^{+0.17}$ & $0.37_{-0.19}^{+0.25}$ \\
$e_{\rm{d}}$ (derived) & $0.28_{-0.16}^{+0.21}$ & $0.21_{-0.11}^{+0.17}$ & $0.50_{-0.20}^{+0.14}$ \\
$\rm{rms}\ (\rm{m\ s^{-1}})$ & $6.72_{-0.49}^{+0.56}$ & $6.60_{-0.44}^{+0.49}$ & $6.72_{-0.48}^{+0.55}$ \\
$\sigma_{\rm{white}} (\rm{m\ s^{-1}})$ & $4.31_{-0.32}^{+0.33}$ & $6.00_{-0.42}^{+0.46}$ & $4.31_{-0.32}^{+0.35}$ \\
$\sigma_{\rm{red}} (\rm{m\ s^{-1}})$ & $5.96_{-0.59}^{+0.66}$ & $3.53_{-0.33}^{+0.38}$ & $5.96_{-0.32}^{+0.35}$\\
$N_{\rm{sim}}$ & 1961 &  1957 & 1941 \\
\hline\hline
\end{tabular}}
\begin{tabnote}
\hangindent6pt\noindent
\hbox to6pt{\footnotemark[$*$]\hss}\unskip%
Injected noises do strictly follow the values in this table, but with another small error, about 10\% at most. $N_{\rm{sim}}$ is the number of adopted simulations for each model.\\
Orbital periods $P$ and semi-amplitudes $K$ are not given in this table since the true values are all within 1$\sigma$ from the simulations (Figure \ref{fig:sim}).
\end{tabnote}
\label{tab:sim}
\end{table*}

\end{document}